\documentclass[12pt]{article}
\usepackage{amsfonts}
\usepackage{epsfig,amsmath,amssymb,graphics,color,calc,cite,psfrag} 
\begin{document}

\title{Jamming percolation and glassy dynamics}
\date{}
\author{
 Cristina Toninelli\thanks{Institut des Hautes {\'E}tudes Scientifiques, Le Bois-Marie 35, Route de Chartres F-91440 Bures-sur-Yvette, FRANCE} and Giulio Biroli\thanks{Service de Physique Th{\'e}orique, CEA/Saclay-Orme des Merisiers,
F-91191 Gif-sur-Yvette Cedex, FRANCE}}

\maketitle
\begin{abstract}
We present a detailed physical analysis of 
the dynamical glass-jamming transition which occurs
for the so called Knight models recently introduced and analyzed 
in a joint work with D.S.Fisher\cite{letterTBF}. Furthermore,
we review some of our previous works on Kinetically Constrained Models.

The Knights models correspond to a new class of kinetically constrained 
models which provide the first example of finite dimensional models with
an {\sl ideal glass-jamming transition}.
This is due to the underlying percolation transition 
of particles which are mutually
blocked by the constraints. This
{\sl jamming percolation} has unconventional features:
it is discontinuous (i.e. the percolating cluster is compact at the transition) and the typical size of the clusters diverges faster than any power law when $\rho\nearrow\rho_c$.
These properties give rise for Knight models to an 
ergodicity breaking transition at $\rho_c$: at and above $\rho_{c}$ a 
finite fraction of the system is frozen.
In turn, this  finite jump in
the density of frozen sites leads to
 a  two step relaxation for
dynamic correlations in the unjammed
phase, analogous to that of glass forming liquids. 
Also, due to the faster than power law divergence 
of the dynamical correlation length,
relaxation times diverge in a way similar to the Vogel-Fulcher
law.

\end{abstract}


\section{Introduction}

The formation of amorphous solids as glasses and granular media, i.e the
{\sl glass and jamming transitions}, are still unsettled and fascinating
questions in condensed matter physics despite all the works  that
have been devoted
to this subject.
These phenomena occur in a variety of
systems which, even if microscopically very different, share common
features.  Among the others we recall supercooled liquids, colloidal
suspensions and non-thermal jamming systems (e.g.  vibrated granular
materials) \cite{liquids,colloids,Dau}. 
 Basic {\sl glassy
  properties} include a dramatic slowing down of dynamics when a
proper external parameter is tuned (e.g.  temperature is lowered for
liquids) and the occurrence of a complicated relaxation: non
exponential and spatially heterogeneous \cite{hetexp}.  When
relaxation times become longer than experimental scales, equilibrium
can no more be achieved and the systems freeze into an amorphous
phase.

Even the basic issues in understanding these phenomena
remain unsolved. In particular, it is not
settled whether the dynamical arrest is due to the proximity of a phase
transition and whether this putative phase transition 
is a static or purely dynamical one.
However experiments make it clear that, if an {\sl ideal glass
transition} occurs, it should have an unconventional behavior with
mixed first and second order features. On the one hand, the divergence of
relaxation times and the fact that both  entropy and internal energy 
seem continuous (for molecular liquids)
is indicative of a second order transition.
On the other hand,  there is
a discontinuous order parameter: the infinite time limit of the
Fourier transform of the density--density correlation (analogous to
Edwards-Anderson parameter for spin glasses) has a finite jump at the
transition. This corresponds to the fact that the modulation of the 
microscopic density profile of the glass does not appear continuously
from the flat liquid profile (we generally refer for all systems
to glass and liquid phase meaning the regime before
and after the freezing into an amorphous solid). 
Besides these mixed first/second order properties,
another unconventional feature, compared to usual phase transitions,
concerns the scaling of relaxation
times. Different functional forms have been proposed in the
literature. A very popular and successful representation is
the Vogel-Fulcher law. This suggests that the {\it logarithm} of the
relaxation time diverges as the inverse of the distance from the
transition, i.e. $\log\tau\simeq 1/T-T_0$, for molecular liquids.
Finally, one of the most puzzling features is the absence of 
any experimental evidence of a
static diverging correlation length.  In particular, the dramatic
slowing down of dynamics does not seem to be due to an increasing long range order as for e.g.
ferromagnetic transition:
typical glass configurations are not very different
from instantaneous configurations of the liquid. 
An enormous amount of theoretical approaches have been proposed in the
last fifty years to describe these physical phenomena. However,
if one takes the {\it a priori assumption} that a real glass or jamming phase transition
takes place at a finite temperature $T_{c}$ (at a density less than the close-packed one for the jamming
transition) then the number
of scenarii reduces drastically. Leaving aside very phenomenological
ones, there remain only two: either a very subtle static transition occurs 
(random first order scenario \cite{KTW}) or the transition is purely
dynamical. 
The first possibility invokes \cite{KTW}
a thermodynamic transition similar to the one occurring for
the so called discontinuous spin glass models, e.g. REM and p-spin models \cite{BoCuKuMe}. 
This is a one step replica symmetry breaking transition with a discontinuous
(Edwards-Anderson) order parameter but no discontinuity 
for energy and entropy. 
This approach has been successful in explaining and predicting many
physical phenomena related to the glass transition \cite{Wolynes}. 
Although mean field models upon which it is based are very well
understood by now, some of the finite dimensional predictions are still
semi-phenomenological and need further theoretical works to be put on
a firm and solid basis, see \cite{BB,Franz,WolynesDzero,Moore} for recent works
in this direction.
The possibility of a purely dynamical glass transition
have been mainly investigated through the so called 
Kinetically Constrained Models (KCM)
(\cite{J}---\cite{TB} and references therein).
 These are stochastic lattice gases based on the ansatz that glass or jamming
transitions are due to effective geometrical constraints on the rearrangements of
the atoms or molecules generated close to the transition. Static
correlations beyond those present in dense liquids are assumed to
play no role. Recently, in a joint work with D.S.Fisher
\cite{letterTBF}, we have shown that indeed some KCMs display a purely 
dynamical transition on finite dimensional lattices with the basic 
features of the previously  described glass-jamming transition.

Note that these two scenarii are not necessarily in contrast because it
might well be that, despite similar behaviors, the glass-jamming
transition in molecular liquids, colloids and granular media are of
different origin. The thermodynamic scenario might apply to the glass
transition of molecular liquids whereas jamming transitions might
correspond to purely dynamic phase transitions.

Here we give an extended explanation of the result in \cite{letterTBF}
and we review some of our previous works (and some others when needed) 
on KCM. We focus in particular
on the different behaviors and tools needed to deal 
with different choices of the stochastic dynamics.
This paper is not intended to be an extensive review 
on KCMs: for all the results previous to our
works we refer to the reviews \cite{J,RS} and references therein,
while some more recent results for kinetically constrained spin models can be found in \cite{GC,WBG,JMS,CMRT}.

The outline of the paper is the following.
In Section \ref{KCM} we introduce the models distinguishing those with Glauber (KCSM) and Kawasaki (KCLG) dynamics and further dividing both classes into cooperative and non cooperative models. In Section \ref{kinds}
 we review the different types of dynamical arrest that occur and motivate the analysis of KCMs from 
a theoretical point of view. In Section 
\ref{noncoop} we present  tools and results for non cooperative models.
In Section \ref{bethe}  we describe the jamming transition that cooperative KCMs display
 on Bethe lattices. In Section \ref{knight} we discuss
the class of cooperative models (Knight models) introduced in \cite{letterTBF,longTB} proving that they display an ideal glass transition. In particular, we show that this transition is related to
a new type of percolation transition, which we call {\sl jamming percolation}.
In this Section we also explain which are the tools one needs in general 
to study cooperative KCMs, using Knight models as an example.
Finally, we present our conclusions in Section \ref{conclu}.

\section{Kinetically Constrained Models (KCM)}
\label{KCM}

KCMs are stochastic lattice gases with hard core exclusion, that is on
each site there can be one or zero particle.  In other words, a
configuration on a lattice $\Lambda$ is given by the set of two-valued
occupation variables on each site $x\in\Lambda$:
$\eta_x=\{1,0\}$. These represent occupied and empty sites (particles
and vacancies), respectively. The dynamics is given by a continuous
time Markov process which consists of a sequence of jumps for models
with conservative (Kawasaki) dynamics and birth/death for models with
non conservative (Glauber) dynamics. The former are also known as
Kinetically Constrained Lattice Gases (KCLG), the latter as
Kinetically Constrained Spin Models (KCSM) or facilitated spin models
($\eta_x=\{1,0\}$ are interpreted as spin up and spin down).
For all the models we consider, dynamics satisfies detailed balance
w.r.t. Bernoulli product measure, $\mu_{\rho}$ , at density $\rho$.
 Thus, there are no static interactions beyond
hard core and an equilibrium transition cannot occur. However, in order
for a move to be allowed, it is not enough to verify the hard core
constraint. Indeed the jump or birth/death rates are non
zero only if the configuration satisfies some additional local
constraints, hence the name {\sl kinetically constrained}. These mimic 
the geometric constraints on the possible rearrangements in
physical systems, which could be at the root of the dynamical
arrest \cite{FA,KA}. For example, in a highly dense liquid,
numerical simulations \cite{Kob} have shown that a molecule is caged
by its neighbors and it cannot move on a substantial distance
unless the neighbors move and the ``cage is opened''.  Experimental
evidence of such {\sl cage effect} for colloidal suspensions and
granular materials have indeed been detected \cite{WW,Dau}. 

Numerical simulations show that, for proper choices of the
constraints, KCM display glassy features including stretched
exponential relaxation, super-Arrhenius slowing down and dynamical
heterogeneities\cite{RS}.  Therefore, despite their simplified and discrete
character they could capture the key ingredients of glassy dynamics,
at least on proper time scales.  Indeed, several works have been
recently devoted to understanding the mechanism which induces
these glassy properties and evaluating the typical time/length scales
involved.  

\subsection{Kinetically Constrained Spin Models (KCSM)}
\label{KCM1}

\begin{figure}
\psfrag{a}[][]{{\LARGE{$a)$}}}
\psfrag{b}[][]{{\LARGE{$b)$}}}
\psfrag{c}[][]{{\LARGE{$c)$}}}
\psfrag{NE}[][]{{\huge{NE}}}
\psfrag{SW}[][]{{\huge{SW}}}
\psfrag{SE}[][]{{\huge{SE}}}
\psfrag{NW}[][]{{\huge{NW}}}
\psfrag{x}[][]{{\LARGE{$x$}}}
\psfrag{y}[][]{{\LARGE{$x$}}}
\psfrag{z}[][]{{\LARGE{$x$}}}
\begin{center}
\resizebox{0.99 \hsize}{!}{\includegraphics*{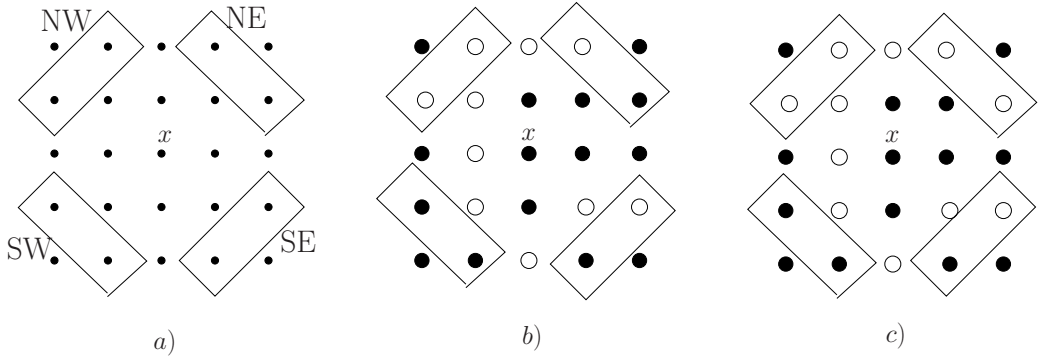}}
\end{center}
\caption{
 a) Site x and the four couples of its North-East (NE),
    South-East (SE),North-West (NW), South-West (SW)neighbours. b) Filled (empty) circles
    stand for filled (empty) sites. Here $x$ is occupied and cannot
    be emptied because in the NorthEast-SouthWest direction none of the two
    couples is completely empty. c) Same configuration as in b) except for
    site $x+2~e_1+e_2$ which is empty. Here $x$ can be emptied since both its
    NW and both its NW neighbours are empty, which guarantees
    $\eta\in{\cal{A}}_x$.}
\label{rule}
\end{figure}

KCSMs are endowed with a non conservative dynamics: each 
site changes from occupied to empty and from empty
to occupied with rate $(1-\rho)f_x(\eta)$ and $\rho f_x(\eta)$,
respectively. The value of $f_x(\eta)$, which
 encodes the kinetic constraints, 
does
not depend on $\eta_x$. Thus detailed balance is satisfied with
Bernoulli product measure, $\mu_{\rho}$, at density $\rho$.\\
KCSMs can be divided into two classes : non-cooperative and
cooperative ones. For the former it is (for the latter it is not) possible
to 
construct an allowed path which completely empties 
{\sl any}
configuration provided a proper finite cluster
of vacancies is present somewhere. We call this cluster a {\sl defect}.
Among non cooperative models we recall the
Fredrickson-Andersen
\cite{FA} one spin facilitated (FA$1f$). For FA$1f$ a
move in $x$ is allowed only if at least one of the nearest neighbors
is empty: $f_x(\eta)=1$ if $~\sum_{y~ n.n. x} (1-\eta_x)>0$,
$f_x(\eta)=0$ otherwise.  This has recently received a
 renewed attention, especially since it has been proposed as a model
 for strong glasses \cite{GCPNAS}. It is easy to check that  the presence of a single vacancy 
in this model allows one to empty the whole lattice.\\
Among cooperative models we recall FA$f$ on an hyper-cubic
lattice of dimension $d$ with $2\leq f\leq d$
\cite{FA}. Here 
the constraint requires that at least $f$ of the surrounding sites
are empty in order for the birth/death rate to be non zero.  As can be
directly checked, for all these models it is not possible to devise a
finite seed of vacancies which allows emptying the whole
lattice. Consider, e.g., the case $d=2$, $f=2$ (with periodic boundary
conditions) and focus on a
configuration which contains two adjacent rows which are completely
filled. These particles can never be erased, even if the rest of the
lattice is completely empty: it does not exist a finite defect which can destroy them, thus the model is cooperative.  The restriction on $f$ comes from the
fact that all cases $f>d$ are very special because at any value of $\rho$ there exist finite sets of
forever blocked particles, namely a fraction of the system is frozen
at all densities. Although they might be interesting in themselves,
they do not seem suitable to describe the slow dynamics close to glass-jamming  
transition because they are non-ergodic at any temperature/density.\\
A different class of cooperative models is the one 
we introduced in \cite{letterTBF}, which we call Knight models.
On a square lattice kinetic constraints are the followings:
a move on $x$ can occur only if [the two
North-East {\sl or} the two South-West fourth nearest neighbors are
empty] {\sl and} [the two North-West {\sl or} the two South-East forth
nearest neighbors are empty]. 
The definition of the North-East, South-East, North-West and South-West neighbors and an example of the constraints is given in Fig.\ref{rule}. 
In formulas $f_x(\eta)=f_x(\eta)^{NE-SW}f_x(\eta)^{NW-SE}$ where
$$f_x(\eta)^{NE-SW}=[(1-\eta_{NE1_x})(1-\eta_{NE2_x})+(1-\eta_{SW1_x})(1-\eta_{SW2_x})]$$
$$f_x(\eta)^{NW-SE}=[(1-\eta_{SE1_x})(1-\eta_{SE2_x})+(1-\eta_{NW1_x})(1-\eta_{NW2_x})]$$
\noindent
with
$NE1_x=x+2e_1+e_2$, $NE2_x=x+e_1+2e_2$, $SW1_x=x-2e_1-e_2$, $SW2_x=x-e_1-2e_2$,
$SE1_x=x+2e_1-e_2$, $SE2_x=x+e_1-2e_2$, $NW1_x=x-2e_1+e_2$ and $NW2_x=x-e_1+2e_2$.\\
Finally, we recall the one-dimensional East model. In this case
the constraint requires a vacancy on the right nearest neighbors, that is $f_x=(1-\eta_{x+1})$. Note that on a finite lattice the presence of
a single vacancy on the rightmost site allows to empty the whole lattice. However the model does not belong to the above defined non-cooperative class, since the vacancy should occur in a specific position.
Indeed, it is usually classified as cooperative:
 due to the direct nature of constraints the relaxation  involves the
cooperative rearrangements of large regions as $\rho\nearrow 1$ \cite{SE,AD,GC}.

\subsection{Kinetically Constrained Lattice Gases (KCLG)}

KCLG are endowed with a conservative dynamics. A particle in $x$ attempts at a fixed rate to jump to a random nearby empty site $y$ and the move occurs with rate $\eta_x(1-\eta_y)f_{x,y}(\eta)$. Here $f_{x,y}(\eta)$ does not depend on the configuration on $x$ and $y$.
Since dynamics preserves the number of particles, detailed balance on finite volume is satisfied w.r.t. the measure which is uniform on configurations with fixed particle number.
On the other hand, on infinite volume detailed balance holds w.r.t 
any Bernoulli product measure $\mu_{\rho}$.
Again, we can classify KCLG as 
non-cooperative and cooperative  models. 
For the former it is (for the latter it is not) possible
to construct a finite group of vacancies such that for {\sl any} configuration it can be moved all over the lattice and any jump of a particle to a neighboring empty site can be performed when the particle is adjacent to the empty cluster.
We will call this mobile cluster which facilitates jumps a {\sl macrovacancy}.
As can be immediately checked
the model in which there are no further kinetic constraints besides 
hard core, the so called normal lattice gas or symmetric simple exclusion process (SSEP),
is non cooperative and the minimal macrovacancies are simple vacancies.

Among cooperative models we recall Kob Andersen model (KA) \cite{KA}
on a cubic lattice. Here a particle can jump to a neighboring site
only if both in the initial and final position at least $m$ of
its nearest neighbors are empty, where $m=3$. Analogously, one can
define KA models on hyper-cubic lattices of dimension $d$ and for
different values of the parameter $m$, with $2\leq m \leq d$. The
restrictions on $m$ are due to the fact that $m=1$ corresponds to SSEP
 while a model with
$m>d$ has a finite fraction of particles 
that is frozen at any density 
(cf. FA models with $f\geq d+1$). Again, one can directly check that all 
these models are cooperatives according to the definition above.
Consider, e.g. KA in $d=3$ with $m=3$ and focus on a configuration
containing a completely filled
slab  which spans the lattice and has a two by two transverse section. 
Any finite cluster of vacancies can never
overcome this filled structure, therefore it is not possible to
construct a finite macrovacancy that moves everywhere in the lattice. 
Another possible choice of cooperative KCLG
are conservative Knight models \cite{letterTBF} defined as follows. A move from
$x$ to $y$ occurs only if the configuration satisfies
the requirement needed to allow the move in $x$ {\sl and} in $y$ for the non conservative Knight  model defined in previous section (plus $y$ 
should be empty).

On the other hand, an example of a non cooperative KCLG is provided by
KA model with $m=2$ on a triangular lattice. Indeed, two neighboring
vacancies form the required mobile macrovacancy as will be further explained in Section \ref{noncoop}.

\section{Glass-jamming transition in KCM}
\label{kinds}
\subsection{Ergodicity breaking and glass-jamming transitions}
Let us take the point of view that glass and jamming transitions observed 
in experiments are not just cross-overs but they do correspond to a 
real thermodynamic or dynamic phase transition. For simplicity we will focus 
on lattice models assuming that the presence of the underlying lattice does not 
change qualitatively the physics.  

\begin{figure}
\begin{center}
\includegraphics[width=9cm]{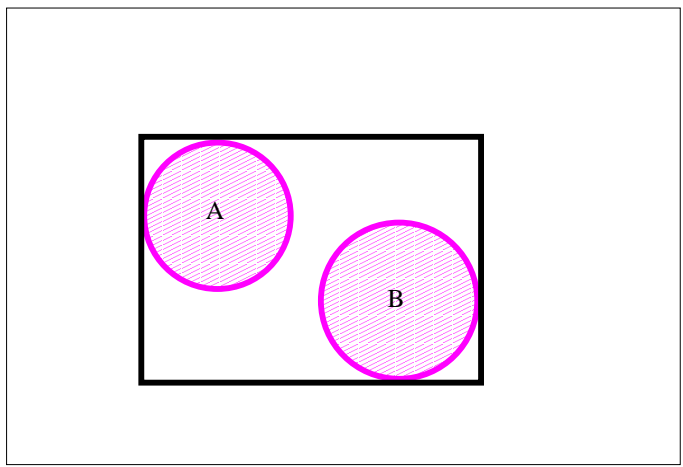}
\end{center}
\caption{The configuration space of the two discs in the box is broken in disconnected
components. It is not possible to reach dynamically the configuration with the disk A 
in the position of B and viceversa.}
  \label{coins} 
\end{figure}

As explained in the Introduction, from the experiments it is clear that {\it ideal glass-jamming transitions} (if they exist) 
have peculiar features compared to standard first or second order phase transitions. 
In particular,  they are characterized by a diverging timescale although 
no growing static correlation lengthscale has ever been found, contrary to 
second order phase transition where the relaxation timescale diverges
because long-range thermodynamic order sets in.  
This strongly suggests that ideal glass-jamming transitions are purely dynamic 
phase transitions, i.e. they are characterized by an ergodicity breaking
without any singularity in the thermodynamics. Is it possible? This question has been addressed in the mathematical 
physical literature (see \cite{spohn} for a detailed discussion) 
focusing on interacting hard core particle systems on finite dimensional lattices. 
If particles interact via a short-range potential and the rates for exchanging a particle with a nearest neighbor
vacancy are always strictly positive then it has been proved that 
in two dimensions the ensemble 
of the Gibbs measures, $\cal G$, coincides with the ensemble of stationary (under the 
dynamical evolution) measures. In higher dimensions it has been proved 
that $\cal G$ coincides with all reversible stationary measures, where reversible measure means 
(roughly speaking) that it verifies detailed balance. It is not known if there    
exist stationary non reversible measures. It seems unlikely because they 
would lead to circulating probability currents without any net injection of     
energy inside the systems, see \cite{spohn}. In any case they are certainly not relevant
for the problem of the glass transition. Thus, as far as glass-jamming transitions 
are concerned, one can safely assume (as it has been proved in two dimensions)
that the ensemble of Gibbs measures coincides  with the ensemble of stationary 
measures. Therefore, if a dynamical transition takes place, this would lead to more than one
stationary measure and as a consequence to more than one Gibbs measure: dynamical 
transitions has to be accompanied by thermodynamic transition. \\
Would this mean that the ideal glass-jamming transitions must be only of thermodynamic
origin and no pure dynamical transition can take place? This is puzzling, because
 the absence of any indication of growing static correlation lengths
seems to be at odds with the very existence of a thermodynamic transition
inducing a diverging relaxation time.
If one does not want to let down the assumption that experimental 
glass-jamming transitions correspond to real phase transitions there remains 
mainly two possibilities. A thermodynamic transition indeed takes place but it is 
of a completely new type and such that is not visible in any given 
$n-$point correlation function. A good candidate for that is the Random First Order 
Theory \cite{Wolynes}.
Otherwise, some hypothesis that lead to the previous conclusion (no purely dynamical transition)
must be violated. The only one that is reasonable to violate on physical grounds
is the fact that the rates 
for exchanging a particle with a nearest neighbor vacancy have to be strictly positive. 
Consider for
example hard spheres with Brownian or Monte Carlo dynamics (that is a reasonably good model for 
colloidal or even granular systems displaying glass-jamming transitions). If the density 
is high enough some moves can have rate zero.
This can have a dramatic consequence on the dynamics because now the configuration 
space can be broken into subsets that are not connected by any dynamical move when  density is large enough, see Fig.\ref{coins} 
for a trivial example.
Formally, 
the Markov chain associated to the dynamical evolution becomes
reducible contrary to the case of the lattice gas with strictly positive rates.\\

As a conclusion the only possibility of having a purely dynamic transition seems to boil
down to having zero (i.e. degenerate) rates depending on the configuration around a site. This
encodes the physical effect explained previously, see also Fig. \ref{coins}, that is certainly present for systems
of hard objects as colloids and granular media. The dynamical transition that may be 
induced by degenerate rates is a reducible-irreducible phase transition. At 
low density (high temperature) almost any equilibrium configuration is contained in 
the same ergodic component. Instead, at high density (low temperature), the configuration 
space sampled with the equilibrium measure is fractured in many different components 
(and no one of them covers almost all the space). 
KCMs are the simplest model that take this phenomenon into account. 
Because of their simplicity they allow to study 
in great detail the issue of glass-jamming transition as purely dynamical transition.

Indeed for KCMs  (see Section \ref{KCM}) some rates are zero, hence their name 
{\it kinetically constrained}, and this implies that
the configuration space on finite lattices is generically reducible. In particular,
there exist configurations which cannot be
connected one to the other using allowed moves. For example,
for FA model with $f=2$ and $d=2$ (see Section \ref{KCM1}),
a configuration which is empty with the exception of
two adjacent rows which are completely filled can never be connected
to any configuration which does not contain this slab. 
This implies that on any finite volume $\Lambda$ the process is
not ergodic and $\mu_{\rho}$s are not the unique invariant measures.
The crucial issue is whether ergodicity is restored in the
thermodynamic limit and whether this depends upon the value of $\rho$.

Two final remarks are in order. First, as shown in \cite{BT,noiKA}, 
thanks to the product form of the equilibrium measure, for KCMs
establishing ergodicity at density $\rho$ corresponds to proving that
there exists an irreducible set of configurations\footnote{More precisely, we should identify a
sequence of sets ${\cal{F}}_L$ 
on ${0,1}^{|\Lambda_L|}$ such that ${\cal{F}}_L$ 
is irreducible w.r.t. the chosen 
dynamics on $\Lambda_L$ with periodic boundary conditions and 
$\lim_{L\to\infty}\mu_{\rho}({\cal{F}}_L)=1$.}  (i.e. configurations connected one to the others by allowed paths) which has unit
probability w.r.t. $\mu_{\rho}$ in the thermodynamic limit.
Another way to put it is that the only possible way of breaking ergodicity 
is by the reducible-irreducible transition discussed above. 
The second remark is that this type of transition is very different from 
the critical slowing down related to second order phase transitions. 
In those cases the ergodicity breaking is induced by the thermodynamic
limit (the number of degrees of freedom goes to infinity and so the relaxation time
may go to infinity). Instead, for the type of dynamical transition advocated
above, finite size systems are typically non ergodic and the crucial question is whether
in the thermodynamic limit the ergodicity is restored and up to which density.

\subsection{Dynamical transition in KCMs $~~~~~~~~~~~~~~~~~~~~~$
and jamming-bootstrap-$~k$core percolation}

For non-cooperative models it is immediate 
to conclude that ergodicity holds 
at any density $\rho<1$. Indeed, the probability of
finding at least a defect or a macrovacancy for an equilibrium configuration 
goes to one in infinite volume. 
Starting from this macrovacancy, by definition of non-cooperative KCSM, 
one can empty all the lattice. Therefore, two typical equilibrium configurations
are almost surely connected by a path (with strictly positive rates) in the configuration space.
The situation is similar for non-cooperative KCLG. In this case
configurations with a macrovacancy can be connected by a path 
which subsequently performs all nearby particle exchanges 
by previously moving the macrovacancy near the interested sites.\\ 
The case of cooperative KCMs is much more involved and indeed it can lead to 
an ergodicity breaking transition. In general, as discussed 
previously, in order to prove ergodicity 
one should construct an irreducible set and prove
that it has unit probability w.r.t. $\mu_{\rho}$ in the thermodynamic
limit.  For KCSMs (with periodic boundary conditions) the
relevant irreducible set is composed by configurations that can be
completely emptied by allowed moves, 
as it can be found easily analyzing these systems at low density. 
This component can be identified
by the following deterministic procedure: subsequently remove all
particles for which the constraint is verified until reaching a
completely empty configuration or one in which there is a backbone of
mutually blocked particles. Note that this backbone is uniquely
determined by the initial configuration: it does not depend on the
chosen order to erase particles. If the backbone is empty, we say that
the configuration is {\sl internally spanned} and, by definition, 
it  belongs  to the relevant irreducible component. 
Therefore, ergodicity breaking takes place if and only if equilibrium 
configurations contain an infinite backbone of blocked sites in the 
thermodynamic limit. Note that,as discussed previously, 
we will consider only choices of the constraints 
such that a blocked cluster has necessarily to be infinite. 

Thus, the problem of the existence of an ergodicity breaking
transition for cooperative KCSMs can be reformulated as a percolation
transition for the final configuration of the above cellular automata.
For FA models the latter coincides with bootstrap percolation
procedure \cite{Adler,AL} or k-core percolation \cite{k-core}: 
a particle is removed if it has less than
$m$ neighbors.  The results in \cite{AL,Sch} establish that bootstrap
percolation converges to a completely empty lattice for all $\rho<1$,
namely no percolation transition occurs. Therefore, it is immediate to
conclude that FA$f$ on any hyper-cubic $d$-dimensional lattice
does not display an ergodicity breaking transition at any $\rho<1$ for
all $f$ and $d$.  On the contrary, for the cellular automata
corresponding to Knight models, a {\it jamming percolation} transition (as we called it)  occurs at a
critical density $\rho_c<1$ as proved in \cite{letterTBF,longTB}.
Therefore, Knight models in infinite volume are ergodic for
$\rho<\rho_c$ and non ergodic for $\rho\geq\rho_c$.  
In Section \ref{knight} we will explain
 the mechanism which induces the percolation
transition for the blocked structures of Knight models and discuss 
its character. \\
For cooperative KCLG, the prove of ergodicity is more involved. For example, for
KA model on square lattices with $m=2$, we have shown that the irreducible component which has unit probability in the thermodynamic limit is the one composed by
configurations which can be connected by an allowed path to a configuration 
which have a frame of vacancies on the boundary \cite{noiKA}. Establishing that
this set has unit probability in the thermodynamic limit is more involved than
in the corresponding KCSM (i.e. FA model). This is due to the fact that
there do not exist a deterministic bootstrap-like procedure
which allows to establish whether a configuration belongs or not
to the irreducible set. 

Finally note that, apart from ergodicity breaking, other types of transition might take
place in lattice models for glass-jamming transitions: (1) 
a change  from an exponential to a 
{\sl stretched exponential relaxation}
 for density-density correlation in KCSM or its
Fourier transform for KCLG, (2) a
{\sl diffusive/subdiffusive transition} in the behavior of a probe particle and (3) a {\sl break down 
of the conventional hydrodynamic limit}.

In the case of cooperative KCSM numerical simulations \cite{KA,GPG,H,F} seems to indicate 
that connected correlation functions, e.g.
$<\eta_x(t)\eta_x(0)>_{\rho}-<\eta_x(t)>_{\rho}<\eta_x(0)>_{\rho}$, decay
in time as stretched exponential, $\exp[-(t/\tau)^{\beta}]$, with a stretched exponent 
$\beta$ that is less than one and density dependent
($\beta$ decreases as $\rho$ is increased).  Analytical works based on mode coupling 
approximations \cite{ES} support these results. 
The standard explanation of this behavior is dynamic heterogeneity, which has
indeed been experimentally detected \cite{hetexp} and consists in having different 
relaxation times in different regions of the system. The superposition of 
these exponential relaxations would then lead to an {\it effective} stretched exponential 
behavior. 

Although this is likely what happens for KCSMs and real liquids, it has been
recently proved in \cite{CMRT} that relaxation is exponential at very large times
for all cooperative and non-cooperative models (in the ergodic regime) considered so far. 
More precisely, relaxation is exponential on times
of the order of the inverse of the spectral gap of the 
Liouvillian operator generating the dynamics (that is 
the slowest relaxation time over all one time quantities)
and this is proved to be finite at any $\rho$ in the ergodic regime. 
The discrepancy between rigorous and numerical results is likely due to the 
different time regimes that are investigated: 
a correlation function can be well approximated as a
stretched exponential 
for times $t$ of the order of its relaxation time 
\footnote{Here we mean the time 
 over which the specific normalized 
connected correlation function has decayed to $e^{-1}$, which may of course be shorter than the slowest relaxation time given by the inverse of the spectral gap.}
 $\tau$ 
and decay as an exponential 
when $t>>\tau$. An interesting question is whether the effective 
numerical behavior coming from fit on times of the order of 
$\tau$ can be made precise: in the scaling limit $\tau\rightarrow\infty$,
with $t/\tau$ of the order of one, do correlation functions converge 
to stretched exponentials?.

Another transition different from the ergodicity breaking is a diffusive/ subdiffusive 
transition for the motion of tagged particles. 
The result that the self-diffusion coefficient, $D_s(\rho)$, 
is positive at any density less than one, 
which is known to hold for SSEP  cannot be
immediately extended to KCLG due to the degeneracy of jump rates.
In particular for KA models, from numerical simulations \cite{KA}
it had been conjectured that $D_s>0$ only below a finite critical density.
However, in \cite{BT} it has been proved $D_s>0$ at any density $\rho<1$ for non-cooperative models and the result have been extended to cooperative models (in the ergodic regime) in \cite{TB}.

Finally, a third possibility is a breakdown  of the conventional hydrodynamic limit that usually holds on long length and time scales \cite{spohn}. For KCLG, if the constraint is released, the macroscopic density profile evolves via the diffusion equation
$\partial_t\rho=\nabla(D(\rho)\nabla\rho)$ where $D(\rho)=1$ for the normal lattice gas. Again, due to the presence of constraints, one cannot apply the techniques which have been developed to establish this hydrodynamic limit for stochastic lattice gases \cite{librolandim}.
Furthermore, for some cooperative KCLG, it has been conjectured \cite{jorgemauro} that the macroscopic diffusion coefficient $D(\rho)$ would vanish at high density, leading to a sub-diffusive evolution of density profiles.
In \cite{LT} it is proved that  for a class of non cooperative models
hydrodynamic limit holds with $D(\rho)$ vanishing (as power law of $1-\rho$)
only for $\rho\to 1$. Extending these results to cooperative models
would require establishing the scaling of relaxation
times on finite lattices with the size of the lattice,
which has not yet been settled for
cooperative KCLG.
 
\section{Non-cooperative models as renormalized lattice gases}
\label{noncoop}

In Section \ref{kinds} we explained that, thanks to the presence of
the macrovacancies (for KCLG) and defects (for KCSM),
 in the thermodynamic limit non
cooperative models are ergodic at any $\rho<1$ as it occurs for the
corresponding models without kinetic constraints. 
In this Section we shall
explain that the slow dynamics for non cooperative models can
be understood in terms of motion of macrovacancies or
defects, i.e. as renormalized lattice gases. 
Indeed these mobile regions, when $\rho\to 1$, essentially perform
independent random walks and substantial relaxation takes place only
when they pass by. So the high density dynamical behavior is encoded
in their properties (size, density, timescale for motion, \dots).
The analysis and the physical scenario is very similar for non-cooperative
KCSM and KCLG. In the following we shall focus on KCLG which are in a sense
richer because, apart from the relation timescale, one can study directly 
the self-diffusion coefficient of tagged particles, $D_s$,bulk diffusion 
coefficients, $D(\rho)$, etc.

Let us first focus on the scaling with density of $D_s$.
Consider for example KA model with $m=2$ on a triangular lattice (see 
Section \ref{KCM}). 
\begin{figure}
\centerline{
\includegraphics[width=0.5\columnwidth]{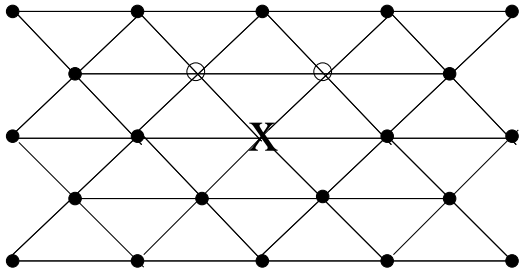}}
\caption{The triangular lattice. For KA with $m=2$ a
 particle in $x$ can be moved to the neighbouring empty 
site $x+(e_1+e_2)/2$ thanks to the fact that
the other empty site ($x+(-e_1+e_2)/2$)
is nearest neighbors of the particle both in its initial and final position.}
\label{tria1}
\end{figure}
The triangular lattice
$\Lambda$, represented in Fig. \ref{tria1}, is the union of sites
in a square lattice $\Lambda_1$ and in its dual $\Lambda_2$ which is
obtained by displacing $\Lambda_1$ of $(e_1+e_2)/2$ with $e_1=(1,0)$,
$e_2=(0,1)$.  Two sites $\{x,y\}\in\Lambda$ are nearest neighbors if
$x-y=\pm e_1$ or $x-y=\pm (e_1+e_2)/2$ or $x-y=\pm (-e_1+e_2)/2$.
Dynamics can be reformulated in terms of
vacancy motion: a vacancy can move to a neighboring
occupied site only if it has at least one empty neighbor
both in its  initial and final position.  Since two neighboring sites of
the triangular lattice share a common third neighbor, it is immediate to see
that any of the two vacancies of a neighboring couple can move to the
common third neighbor. Therefore a couple of neighboring
vacancies is a finite size freely mobile cluster. Furthermore 
 it is possible to move any given particle into an
empty nearest neighbor provided the couple of vacancies is nearby.
For
example, if we want to move a particle from $x$ to $x+(e_1+e_2)/2$ it is sufficient
to put the macrovacancy in $x+(e_1+e_2)/2, ~x+(-e_1+e_2)/2$, as shown in Fig.\ref{tria1}. Thus, a couple of neighboring vacancies is a macrovacancy according to the definition in Section \ref{KCM}.
A simple heuristic argument based on the independent motion of these macrovacancies
leads to the correct high density dependence of $D_s$.
Let us focus on dimensions larger than two (for lower dimension the 
reasoning changes due to recurrent properties of random walks).
Call $\rho_{d}$ the density of macrovacancies
and $\tau_{d}$ the timescale on which they move. The self-diffusion
coefficient is expected to be proportional to the
inverse of the time $\tau_{p}$ on which each particle moves of one
step. On the timescale $\tau_{d}$ the number of particles that have
jumped is of the order $V\rho_{d}$  where V is the total number of
sites. Thus we find
\begin{equation}\label{}
\frac{\tau_{p}}{\tau_{d}}V\rho_{d}\propto V\rho.
\end{equation}
As a consequence, at density close to one, we get $D_{S}\propto
\rho_{d}/\tau_{d}$.  Since for the $m=2$ KA on a triangular lattice 
macrovacancies are formed by two
neighboring vacancies, in the limit $\rho
\rightarrow 1$ we get $\rho_{d}\propto (1-\rho )^{2}$ and $\tau_{d}\propto O
(1)$. Hence, $D_{S}\propto (1-\rho )^{2}$. The same arguments
leads to $D_{S}\propto (1-\rho)$ for SSEP,
where macrovacancies are single vacancies. The result for SSEP has been rigorously 
proved in \cite{S} by establishing upper and lower bounds.
For KA model on a triangular lattice, as we have shown in \cite{TB} (see also \cite{BT} for a different choice
of non cooperative constraints), it is also possible to turn the heuristic argument
 into a proof deriving upper and lower bounds
$c_l(1-\rho)^2\leq D_s(\rho)\leq c_u (1-\rho)^2$ with $c_u$ and $c_l$ independent
from $\rho$.

The representation of non-cooperative KCM as renormalized lattice gases 
turns out to be useful also to obtain these rigorous results.
Let us recall the basic strategy since it allows to further 
understand the role
of macrovacancies.  
The
idea is to start from the proof for SSEP in \cite{S} 
and to modify it by letting macrovacancies play the role
of vacancies.  The procedure is the following: first construct a suitable 
auxiliary model that is easy to analyze and then, using a variational
formula for $D_s$, show that $D_s>h(\rho)D^{aux}_s$ where $h(\rho)$ is a 
function determined explicitly (that is strictly positive for $\rho<1$). 
The technique for the upper bound is more classical \cite{S} and the lower bound 
is more important because it shows that, despite the dynamics slows down, particles 
still diffuse on large distances and timescales. Therefore, we will just focus 
on the lower bound in the following.

In \cite{S} the auxiliary model is constructed in this way: 
the tracer has a neighboring vacancy at time zero and
the only possible moves are jumps of the tracer and exchanges of the
occupation variables in $y$ and $w$, where both $y$ and $w$ are
nearest neighbors of the tracer.  Note that the latter moves 
are not allowed for SSEP but one can reconstruct them 
trough a path of allowed moves thanks to the presence 
of the vacancy. For the auxiliary model it is
immediate to prove that $D_s^{aux}>0$ and is bounded from above and below by density independent constants. Indeed, a  move of the tracer from $x$ to $x+e_i$ 
can always occur via two steps: since the tracer has always at least one neighboring vacancy, this can be brought
 (in one move) in $x+e_1$ and then the jump of the tracer from $x$ to $x+e_1$ can occur.
Using a variational formula for $D_s$ \cite{S} and 
comparing the dynamics of the normal lattice gas and the auxiliary model
leads to $D_s^{SSEP}> c~(1-\rho) ~D_s^{aux}$. The term 
$(1-\rho)$ comes from the requirement of having at least a
vacancy at time zero near the tracer; $c$ is a density independent constant which
accounts for  the maximal length of the path needed to
 exchange the occupation variable in $x$ and $w$ trough
neighboring jumps (the only moves allowed for SSEP).
 For KA on the triangular lattice
one defines the following auxiliary process. 
We require  a macrovancancy near the tracer in the initial configuration
and we allow only moves which either displace the tracer to a nearest neighbor or
move the macrovacancy into another couple of sites which are also  near the tracer
(for a precise definition of the auxiliary model see \cite{TB}). Again, the latter 
move is not allowed in KA but can be reconstructed by a finite path of allowed moves.
As before, it is easy to show that $D_s^{aux}>0$ and is independent from
$\rho$. Comparing KA and  auxiliary dynamics we get $D_s>c~(1-\rho)^2$, where $(1-\rho)^2$ comes from the requirement of having a
macrovacancy.  Note that this procedure
is generalizable to all non cooperative models: if the macrovacancy is formed by $q$ empty
sites it leads to
$D_s(\rho)\geq c (1-\rho)^q$.
Therefore, at least for the motion of the tracer,
non cooperative KCLG are simply
renormalized SSEP: the typical diffusion time of the tagged particle (i.e. the
 inverse of $D_s$) goes here as the inverse of the
density of macrovacancies, instead of vacancies.

The strategy of comparing dynamics with the one of a faster unconstrained model and 
reconstructing the moves of the latter by a proper path of allowed moves
can be also used to derive upper bounds for the density and size dependence of the relaxation time, $\tau$, for 
non-cooperative KCLG and KCSM (see 
\cite{BT} and \cite{CMRT}, respectively).
The paths are always constructed using the fact that, if there is 
a macrovancancy (defect),
we can move it everywhere and facilitate
 any nearest neighbor
jump (birth/death) of particles.
Therefore comparison with the unconstrained model in general gives an extra factor
$(1-\rho)^q$ (the cost of creating the mobile region) times a term related to the 
length of the path (which can bring a dependence both on the lattice size, $L$, and on $\rho$). 
Note that in principle also an entropy term should be accounted (for the bounds on $\tau$ and $D_s$), counting 
all the possible configurations which have to pass trough the same bottleneck \cite{noiKA,BT,CMRT}. 
However, this is just a constant factor for non cooperative models (more precisely it is
the entropy on a number of sites proportional to the size of the macrovacancy)
and cannot change the scaling in $1-\rho$.\\
From results in \cite{BT,CMRT}, one has that for non-cooperative models
the dependence of $\tau$ 
on the lattice size $L$ is the same as for models without constraints:
$\tau\propto L^2$ for KCLG, $\tau$ bounded by a finite
constant uniformly in $L$ for KCSM (where
$\tau$ is the inverse of 
the spectral gap of the Lioviullian operator, see \cite{spohn}).
On the other hand, since relaxation occurs trough the macrovacancies
which become rarer at higher density, the density dependence is different.
For example, for KA on a 
triangular lattice with particle sources at the boundary, we find
$L^2/(1-\rho) c_l<\tau<L^2/(1-\rho)^2c_u$, to be compared with 
$c_l~L^2<\tau<c_u~L^2$ for SSEP.

\section{Jamming percolation on Bethe lattices}\label{bethe}

Bethe lattices (or in the mathematical literature  random c-regular
graphs) are often used in the physics literature as an
approximation of finite dimensional, e.g. hyper-cubic, lattices. 
Because locally they have a Cayley tree structure with connectivity
$k+1$, their analysis is greatly simplified and one can
often obtain exact results on the thermodynamics and the dynamics of
the model embedded on them. 
For proper choices of the constraints, KCMs display a jamming transition on Bethe lattice \cite{Sellitto,RJM} 
and, thus, provide an almost solvable example of jamming transition. 
In the following we present a summary of the results that have been
obtained focusing on a simple case: the 
f-facilitated Fredrickson-Andersen model on a Bethe lattice with 
connectivity $k$ (we consider $k\geq f>1$ in order to avoid any finite
blocked cluster). The results remain qualitatively the same for 
particle models but they are more tedious to derive \cite{noiKA}.

As discussed previously, a jamming (and ergodic) transition for FA
models will take place if and only if at a certain density an infinite
cluster of blocked (jammed) particles appears. All the particles inside this
cluster must have at least $f$ blocked neighbors in order to be
blocked themselves. Thus, this infinite cluster is the spanning
cluster of the so called $f-$core percolation \cite{k-core}, (or
bootstrap percolation \cite{ChLeRe}) a problem that has received a lot
of attention recently \cite{Mendes,Schwarz,Montanari}.
  
\begin{figure}[bt] 
\begin{center}
\includegraphics[width=9cm]{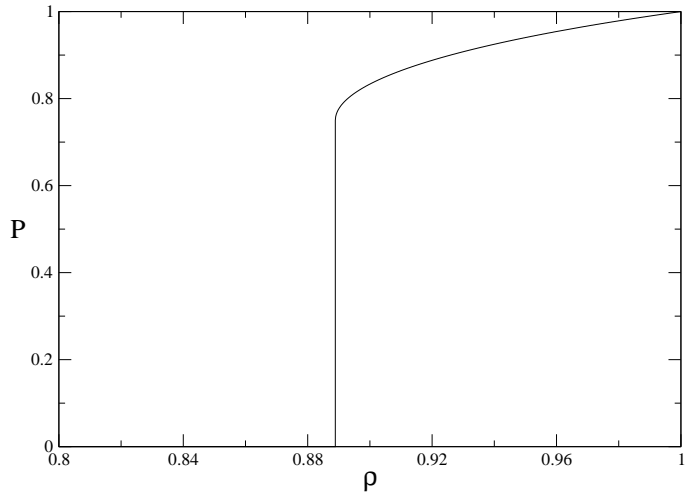}
\end{center}
\caption{Solution of the equation (\ref{f-coreeq}) as a function of
$\rho $ for $k=3$ and $f=2$.}
  \label{figp} 
\end{figure}

For a given site the Bethe lattice local structure is Cayley-tree like
with $k$ branches going up from each node and one going down. Using
this crucial feature it is easy to write a self-consistent equation on
the probability $P$ that  a site is occupied by a blocked particle 
(belongs to the $f-$core) because it has more than $k-f+1$ neighboring
particles above it which are blocked (without taking advantage of
vacancies below)\cite{ChLeRe}:
\begin{equation}\label{f-coreeq}
P = \rho \sum_{i=0}^{f-1} {k \choose i} P^{k-i} (1-P)^{i} .
\end{equation}
Except for $k=f$ for which the transition is continuous, the solution
of this equation leads to $P=0$ for $\rho<\rho_{c}$ and $P=P_{c}+O
(\sqrt{\rho-\rho_{c}})$ with $P_{c}>0$, see the example for $k=3$ and $f=2$
in Fig. \ref{figp}.
 The number of blocked particles or the number of
sites belonging to the $f-$core are polynomial functions of $P$ that
can be derived easily and that have the same behavior of $P$. They are
at the same time discontinuous as in first order phase transitions and
singular because of the square root, as in second order phase
transitions.

The mechanism behind this behavior has been understood in detail
\cite{Mendes,Schwarz}. In particular the singular square root behavior
is due to the extreme fragility of the infinite spanning jammed
cluster at the transition and to the existence of a related diverging
lengthscale. In fact, close to the transition,
 a given jammed particle is
connected to clusters of jammed particles that are only marginally
blocked, i.e. such that removing only one of their blocked neighbors
is enough to unblock them.  These clusters form the so called {\it
corona} of the infinite jammed cluster (or $f-$core spanning
cluster). The size distribution of corona clusters can be computed
analytically \cite{Mendes} and it has been found that their average
size diverges at the jamming ($f-$core percolation) transition as
$1/\sqrt{\rho-\rho_c}$ coming from the jammed phase. 
This explains the singular square root behavior
found previously\cite{Mendes}: 
roughly speaking, decreasing the temperature from
$\rho_c+\epsilon$ to, say, $\rho_c+\epsilon/2$ (where $0<\epsilon<<1$) one
unblocks first a number of particles proportional to $N\epsilon$. However,
unblocking a particle unblocks also all the marginally blocked (corona)
particles attached to it. This leads to a ``domino effect'' such that the
effective number of unblocked particles is $N\epsilon \times
1/\sqrt{\epsilon}$, i.e. the net change in the fraction of blocked
site is $\sqrt{\rho-\rho_c}$. Since corona clusters, i.e. marginally stable
clusters, percolate at the transition one expects an associated
diverging lengthscale. This is indeed the case, see
\cite{Mendes,Schwarz}: coming from the jammed phase there is
lengthscale that diverges as $(\rho-\rho_c)^{-1/4}$.

The analysis of the jamming transition coming from the unjammed phase
is much more difficult and one has to resort to numerical
simulations. In \cite{Sellitto,Berthier} the persistence, 
\[
P (t)=\frac{1}{N}\sum_{i}\langle \Omega_{i} (t)\rangle 
\]
where $\Omega_{i} (t)$ equals one if the site $i$ has remained in the same state (empty or occupied) 
from time $0$ to time $t$ and zero otherwise, and the occupation variable correlation function, 
\[
C (t)=\frac{1}{N}\sum_{i}\langle \eta_{i} (t)\eta_{i}(0)\rangle 
\]
have been measured starting from equilibrium initial conditions.  
\begin{figure}
\begin{center}
\includegraphics[width=9cm]{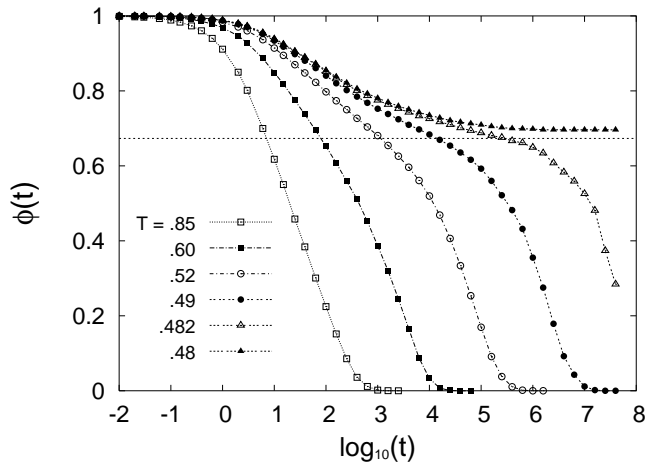}
\end{center}
\caption{Persistence as a function of time for the $f=2$, $k=3$ case  for different temperatures approaching the critical temperature $T_c\simeq0.48$.}
\label{figk3f2}
\end{figure}
We plot in Fig. \ref{figk3f2} the persistence as a
function of time for the $f=2$, $k=3$ case. Different curves 
corresponds to different densities approaching the critical point 
\footnote{Note that in the work  \cite{Sellitto}
we used a different terminology: spin instead of occupation variables and 
temperature instead of density. The relationship between density and temperature $T$ is $\rho=1/(1+e^{-1/T})$.}.
The correlation has a similar behavior \cite{Berthier}. The existence 
of the plateau is a direct consequence of the discontinuous character 
of the jamming transition: the fraction of blocked particles is strictly
positive at the transition coming from the jammed phase instead it is
zero coming from the unjammed one. This translates into a dynamical 
behavior such that $\lim_{\rho\nearrow \rho_{c}}\lim_{t\rightarrow
\infty }P (t)=0$ and $\lim_{t\rightarrow
\infty }\lim_{\rho\nearrow \rho_{c}}P (t)>0$ and equal to the
fraction of blocked sites. Associated to the non-commutation of these
two limits there is a diverging time scale, the time on which $P (t),C
(t)$ equals, say, half of their plateau value. The numerics
indicates that it follows a power law divergence $\tau \propto
(\rho_{c}-\rho)^{-\gamma }$ with $\gamma \simeq 2.9$.

\begin{figure}
\begin{center}
\includegraphics[width=9cm]{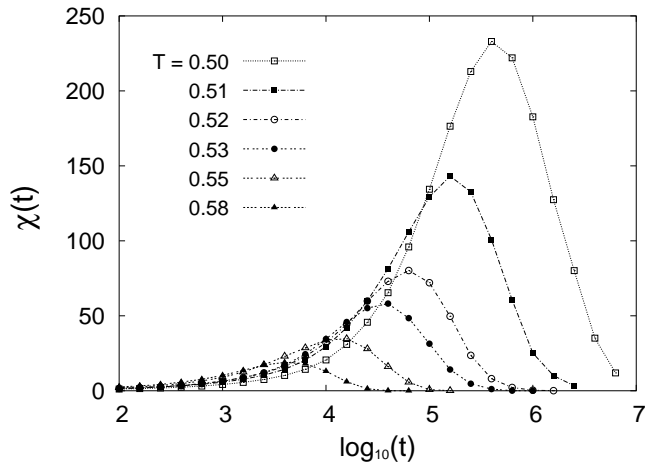}
\end{center}
\caption{Equilibrium dynamical susceptibility $\chi(t)$ vs time $t$
  at temperature~$T$. System size $N=2^{14}$.}
\label{fig_chi}
\end{figure}

Finally, it has been found a diverging dynamical lengthscale associated with the 
divergence of the timescale \cite{Sellitto} (let us recall that no
static correlation can be found because the equilibrium measure is
uncorrelated from site to site). In order to unveil this length one has to focus on
the fluctuations of the persistence or the correlation\cite{FrPa}. 
These are encoded in the dynamical susceptibility 
\[
\chi(t) =N\langle \left[\frac{1}{N}\sum_{i}\Omega_{i} (t)
-\frac{1}{N}\sum_{i}\langle\Omega_{i} (t)\rangle \right] ^{2}\rangle/T 
\]
$\chi(t)$ develops a peak (see 
Fig.\ref{fig_chi}
) for $t\propto \tau
$ that diverges for $T\nearrow T_{c}$. 
This divergence 
is related to the fact that close to $T_{c}$ more and more particles have
to evolve in a correlated way in order to make the system relax. 
Note that, exactly as for ferromagnets the divergence of the 
fluctuations of the magnetization is a signature of a diverging
lengthscale, so it is the divergence of the peak of $\chi(t)$ at the
jamming transition. However, the precise characterization of this dynamic
lengthscale and its relationship with the results on the jammed phase 
have still to be worked out.

Let us summarize the features of the jamming transition found on the
Bethe lattice. It has a {\it first order} character because the 
fraction of jammed particles and the long-time limit of correlation
and persistence functions (called in spin-glasses the Edwards-Anderson
parameter) are discontinuous at the transition. At the same time it
is characterized, as {\it second order} phase transitions, by a diverging
timescale and diverging lengthscales. Although the detailed connection
between time and lengthscales (from the jammed and unjammed phase) has
still to be worked out from numerics it seems clear that all these
quantities diverge as power law in $|\rho-\rho_{c}|$. It is interesting to
note that qualitatively these features are also shared by mean-field
disordered systems, as p-spins models, close to their dynamical
transition and the Mode Coupling Theory of the glass transition \cite{BoCuKuMe}. 
The following section is devoted to the analysis of jamming
transitions in finite dimensional models. As we will see some of the
features found on the Bethe lattice persist, for example the 
mixed (first and
second order)  character  of the transition. Instead others, as the
divergence of the relaxation times, change and actually become closer
to what expected for glass-jamming transitions: 
now the logarithm of $\tau $ diverges as a power law in $|\rho-\rho_{c}|$, i.e. in a Vogel-Fulcher like form. 

\section{Jamming percolation on finite dimensional lattices}
\label{knight}

In this Section we give an extended explanation of the results
obtained in collaboration with D.S.Fisher \cite{letterTBF,longTB} for
the Knight models defined in Section \ref{KCM}.  As already mentioned
an {\sl ergodicity breaking transition} occurs at $\rho_c<1$: above
$\rho_c$ a finite fraction of the system is frozen. This 
is due to an underlying percolation transition for the clusters 
of mutually blocked
particles, which we called {\sl jamming percolation}. 

Let us give a brief summary of the main results letting for the next sections
their derivation.

 As discussed 
in Section 3 the dynamical transition of the Knight models
can be studied using the following 
cellular automata \cite{longTB}. Start from an initial 
configuration sampled with equilibrium measure,
$\mu_{\rho}$, then empty an occupied site if the constraint is
satisfied.  Then continue the
 procedure  until no more particles can
be removed.
 This final configuration is either completely empty 
or one
that contains a percolating cluster of particles which do not satisfy
the constraint. 
Let us call $\rho_{\infty}$ the density of this final
configuration. We will show that at the critical density of site directed
percolation on a square lattice, $\rho_{dp}\simeq 0.705$, a
discontinuous percolation transition occurs: $\rho_{\infty}=0$ for
$\rho<\rho_{dp}$ and $\rho_{\infty}>0$ for $\rho\geq\rho_{dp}$. 
Since the final backbone for this cellular automata contains all the
particles that are frozen under the stochastic evolution of Knights,
we conclude that below $\rho_{dp}$ Knight model is ergodic and
ergodicity is broken above. Furthermore, the fraction of the system
which is frozen coincides with $\rho_{\infty}$ and has a finite 
jump at the transition.\\
As for the FA on a Bethe lattice, we find that 
the dynamical transition transition has a 
{\sl first order} character with a discontinuous
Edwards-Anderson parameter and, at the same time, is characterized by 
diverging time and lengthscales, as second order like transitions.\\ 
The diverging lengthscale can be determined studying finite size effects.
Consider the model on a finite lattice $\Lambda_L$ with
periodic boundary conditions and evaluate the probability $R(L,\rho)$
that it can be completely emptied by allowed moves, i.e. the probability that  $\Lambda_L$ is
{\sl internally spanned}.
Since at any fixed density $\rho<\rho_{dp}$
the final backbone of the cellular automata is empty, $R(L,\rho)$
goes to one in the thermodynamic limit. 
On the other hand, if the limit
$\rho\nearrow\rho_{dp}$ is taken first, $R(L,\rho)$ goes to zero.  The
relevant length $\Xi(\rho)$ is the crossover length which separates
the two regimes: $\lim_{L\to\infty,
\rho\nearrow\rho_{dp}}R(L,\rho)=0$ for $L/\Xi(\rho)\rightarrow 0$ and
$\lim_{L\to\infty, \rho\nearrow\rho_{dp}}R(L,\rho)=1$ for
$L/\Xi(\rho)\rightarrow \infty$.  In other words, $\Xi(\rho)$ is the
dynamical length which corresponds to the
typical size of blocked clusters (i.e. the incipient percolating clusters 
for jamming percolation). As we will explain, by analytical arguments
we find \cite{letterTBF} that
$\log~ \Xi(\rho)\simeq (\rho-\rho_{dp})^{-\mu}$. Here $\mu=\nu_{\parallel}(1-z)$ where $\nu_{\parallel}\simeq 1.73$ is the critical exponent of the parallel directed percolation (DP) correlation length, $\xi_{\parallel}\simeq (\rho-\rho_{dp})^{-\nu_{\parallel}}$, and  $z\simeq 0.63$ is the exponent relating parallel and transverse DP correlation lengths,
$\xi_{\perp}\simeq \xi_{\parallel}^z$.
We also rigorously proved \footnote{The 
existence of the two different correlation lengths, which are due to
the asymmetry of DP and  the power law divergence of $\xi_{\parallel}$, 
are given for granted in physical literature. Indeed, they have
been verified both by numerical simulations and analytical works
trough renormalization technique (see \cite{Hin} for a review). 
However, a rigorous mathematical proof of these results for DP is still lacking. Our  bounds for $\Xi$ are rigorous modulo the
assumption $\xi_{\perp}=\xi_{\parallel}^z$, $z<1$.}
the upper and lower bounds
$c_1\exp({c_2\xi_{\parallel}^{1-z}})<\Xi(\rho)<c_3\exp({c_4 \xi_{\parallel}^{2}})$ \cite{longTB}. These establish that
$\Xi$ diverges faster than power law and the lower bound coincides
with the value we expect for $\Xi$.
\\
Due to this diverging length, 
the relaxation time $\tau$ should diverge as $\Xi^z$ with $z\geq 2$.
Indeed, relaxation should require
the diffusion of regions of density $\Xi$ (see \cite{noiKA}
for similar results for FA model). In \cite{CMRT}
the rigorous bound $\tau\geq \Xi$ is proved, where $\tau$ is the inverse of the spectral gap of the 
Lioviullian operator $\cal{L}$ (i.e. the worst relaxation time on all one time quantities). Therefore Knight models exhibit
a Vogel-Fulcher like relaxation: the log of times diverge as power law when the critical density is approached. This is different from the findings for FA model on Bethe lattices, were $\tau$ diverges as power law (see Section \ref{bethe}).
Since in the Knights model there are no static interactions between the 
particles it is clear that their glass-jamming transition is purely dynamical
and not related to any thermodynamic transition.

In the following we shall explain the arguments leading to the proof of an ergodicity
breaking transition at $\rho_{dp}$ and of its mixed first/second order
character (we refer to \cite{longTB} for rigorous proofs). 
These will unveil
the mechanism which induces these remarkable properties which are 
typical of an ideal glass transition. Also, it will be clear why 
we choose the specific form  of Knight constraints and how one can
modify or generalize them to higher dimension keeping a transition
with the same character.

\subsection{Ergodicity breaking for $\rho>\rho_{dp}$}
\label{noergo}

Let us prove that ergodicity is broken above $\rho_{dp}$,
namely a finite fraction of
sites is frozen. 
This is equivalent to showing
that the origin belongs with finite probability to a percolating
cluster of frozen sites. 
Here, as in
\cite{letterTBF}, we call a site {\sl frozen} if it cannot be
unblocked even by first emptying with allowed moves an arbitrarily
large number of sites.  
\begin{figure}
\centerline{
\psfrag{a}[][]{a)}
\psfrag{b}[][]{b)}
\psfrag{c}[][]{c)}
\includegraphics[width=0.99\columnwidth]{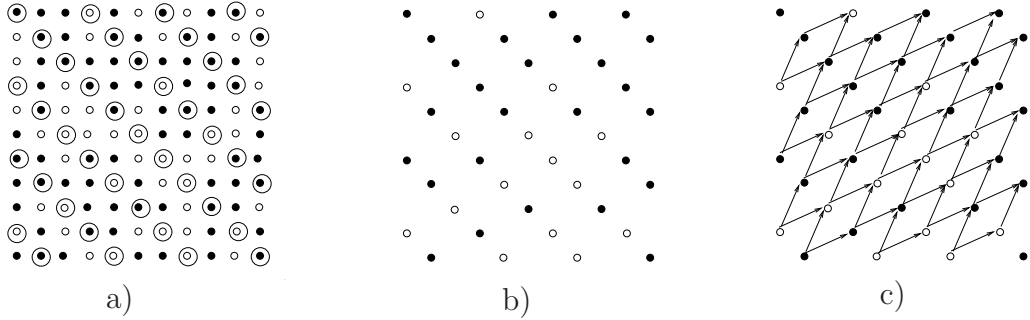}}
\caption{a) Sites inside circles are those 
obtained starting form the origin
and choosing its NE and SW neighbours and the NE and SW neighbours of the latter and so on. b) Sublattice obtained erasing all sites from the square lattice except those inside circles c) Oriented sublattice: arrows go from each site to its two NE neighbours (w.r.t. the original lattice)}
\label{lattice}
\end{figure}
Consider the sublattice which is obtained
from the square lattice
by erasing all sites except the origin, its North-East and South-West
neighbors, the North-East and South-West neighbors of the latter and
so on as in Fig. \ref{lattice} a), b). Then construct a directed
graph on this sublattice by drawing arrows connecting each site to its
North-East neighbors as in Fig. \ref{lattice} c). Notice that this
corresponds simply to a two dimensional square lattice (only rotated and
squeezed).  Also, since there are no static correlations in the
equilibrium measure, if the configuration on the original lattice is
chosen with $\mu_{\rho}$, the same holds for the configuration on the
sublattice. Therefore the results for site Directed Percolation 
\cite{Hin} imply that,  if $\rho>\rho_{dp}$, the
origin belongs with finite probability to an infinite directed
(i.e. following the direction of the arrows) percolating cluster of
occupied sites. It is now easy to check that
if we restore the whole lattice and consider Knight dynamics, all
sites belonging to the cluster which percolates on the sublattice are
frozen: {\sl a finite fraction of the system is frozen above $\rho_{dp}$}. Indeed, each of these sites has at least one occupied NE neighbor
and at least one occupied SW neighbour, therefore it is blocked along
NE-SW diagonal.  In the following, we call
 {\sl NE-SW cluster} a directed occupied path on the sublattice
obtained with the procedure above.  The definition is analogous for
the {\sl NW-SE cluster}, where the sublattice is the one constructed erasing
all sites but the origin, its NW and SE neighbours and so on.  
Note that in
order to establish the existence of frozen clusters we have not used
the possibility of blocking along the NW-SE diagonal. On the other
hand, this should be taken into account if we want to show
that a given configuration does not contain frozen sites as we will
do in the following section.

\subsection{Ergodicity for  $\rho<\rho_{dp}$}
\label{ergo}

\begin{figure}
\centerline{
\psfrag{a}[][]{a)}
\psfrag{c}[][]{b)}
\includegraphics[width=0.45\columnwidth]{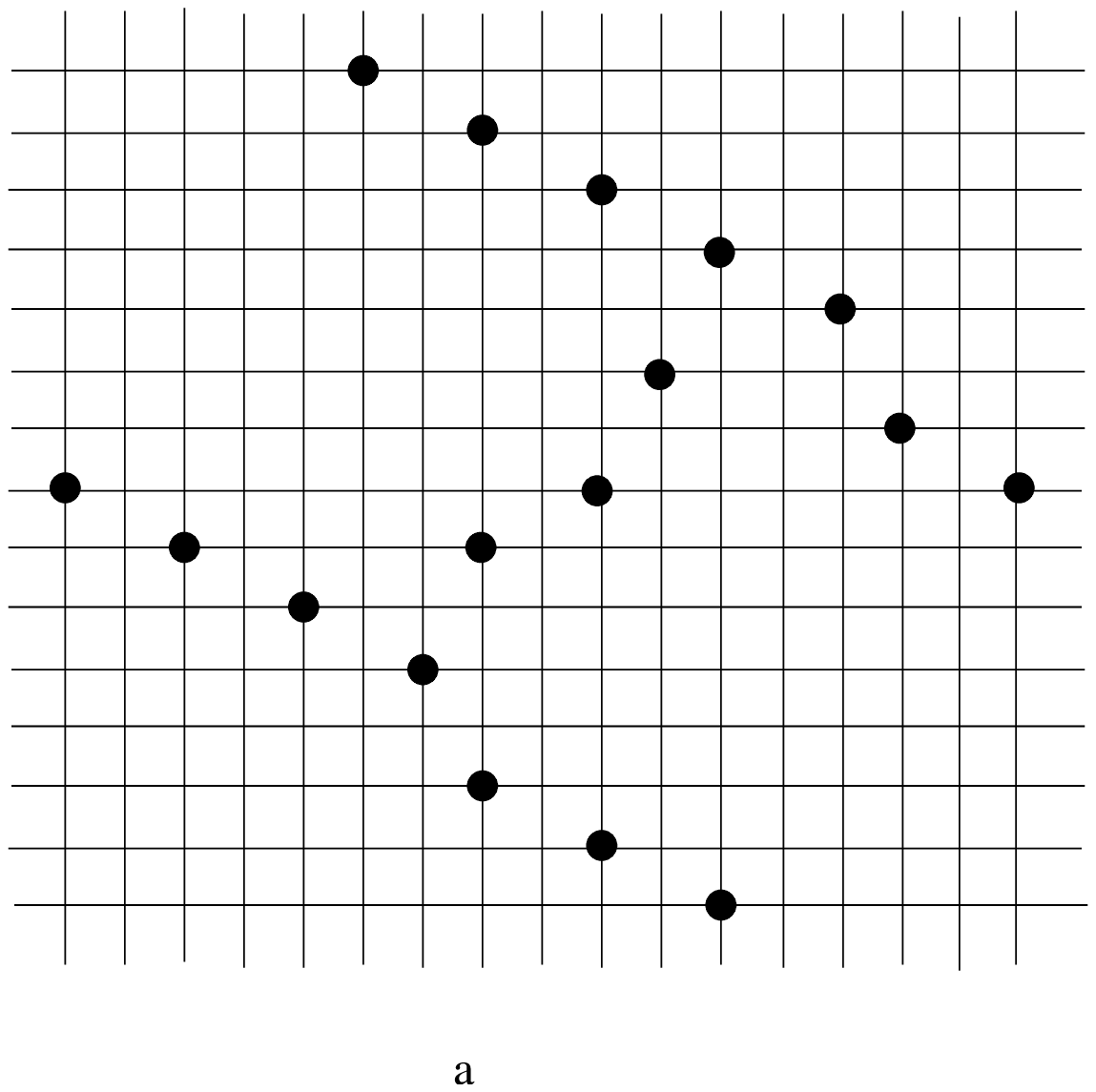}
\hspace{3 cm}
\includegraphics[width=0.45\columnwidth]{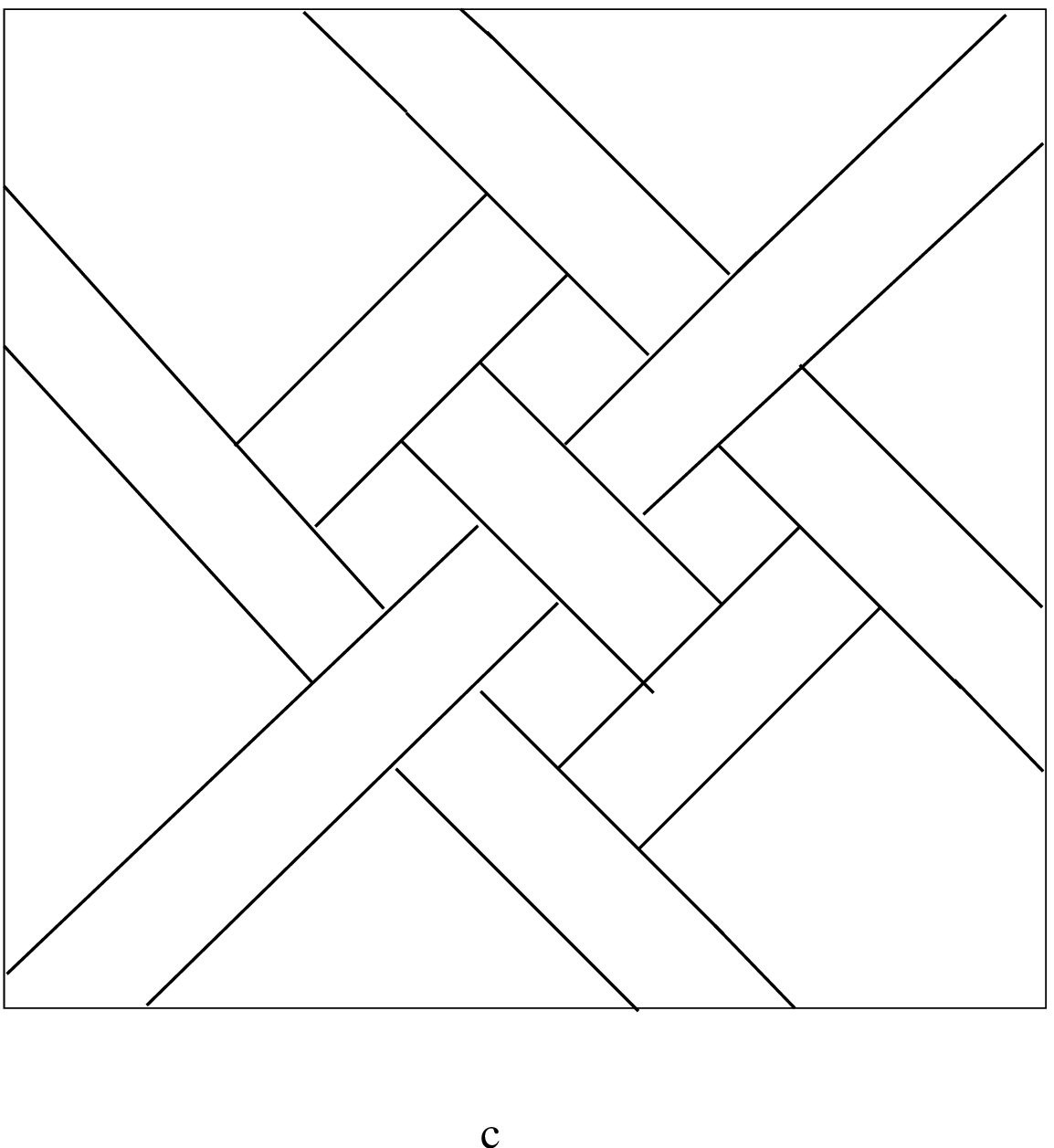}
}\caption{a) A NE-SW non spanning cluster blocked by T-junctions with
two NW-SE spanning clusters; b) Segments parallel to the NE-SW (NW-SE) diagonal
stand for NE-SW (NW-SE) clusters. The depicted structure is blocked,
even if it does not contain neither a NE-SW nor a NW-SE spanning
cluster}
\label{block1}
\end{figure}

Let us prove that the system is ergodic in the thermodynamic limit for
$\rho<\rho_{dp}$, which corresponds to showing that 
the fraction of frozen sites is zero.
Note that if the rule contained blocking only along one of the two
diagonals, ergodicity would follow immediately from the fact that
occupied directed paths do no percolate below $\rho_{dp}$.  
However, for the constraints we have chosen, a percolating directed path
implies a frozen cluster but the converse is not
true. Indeed, due to the fact that a site can be blocked along
either the
NE-SW or the NW-SE diagonal or both, a NE-SW cluster can be blocked
either if it spans the lattice or if it is finite but both its ends
are blocked  by a T-junction with NW-SE percolating paths (see Fig.
\ref{block1} a)). By using such T-junctions it is also possible to
construct frozen clusters which do not contain neither a
spanning NE-SW nor a spanning NW-SE cluster: all
NE-SW (and NW-SE) clusters are finite and are blocked at both ends by T
junctions by finite NW-SE (NE-SW) ones (see figure \ref{block1} b)).
As we will show in detail, these T-junctions are crucial
to make the behavior of jamming percolation very different from 
site directed percolation although the two transitions share the same critical density.

Let us now show how the proof or ergodicity below $\rho_{dp}$ works.
The strategy consists in constructing a set of configurations, ${\cal{F}}_L$,  on $\Lambda_L$ 
and show that ${\cal{F}}_L$ are internally spanned and 
$\lim_{L\to\infty}\mu_{\rho}({\cal{F}}_L)=1$. 
This, as explained in Section \ref{kinds}, 
concludes the proof of ergodicity for $\rho<\rho_{dp}$
and (together with results in previous section) establish
that ergodicity breaking occurs at $\rho_{dp}$.

\begin{figure}
\centerline{
\psfrag{x}[][]{x}
\psfrag{q}[][]{n}
\psfrag{a}[][]{a)}
\psfrag{b}[][]{b)}
\psfrag{m}[][]{n}
\psfrag{n}[][]{I(n/4)}
\includegraphics[width=0.45\columnwidth]{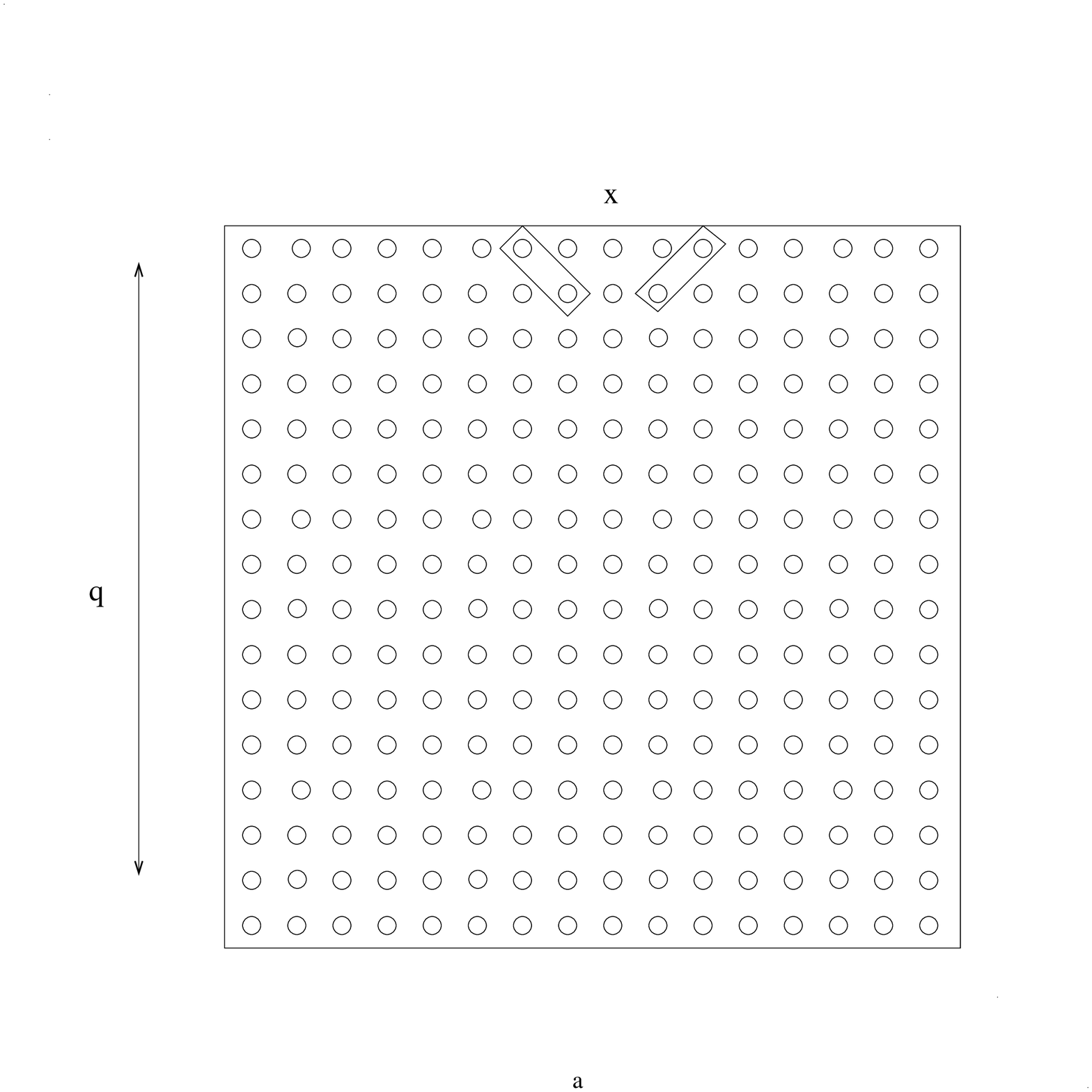}
\hspace{1 cm}
\includegraphics[width=0.45\columnwidth]{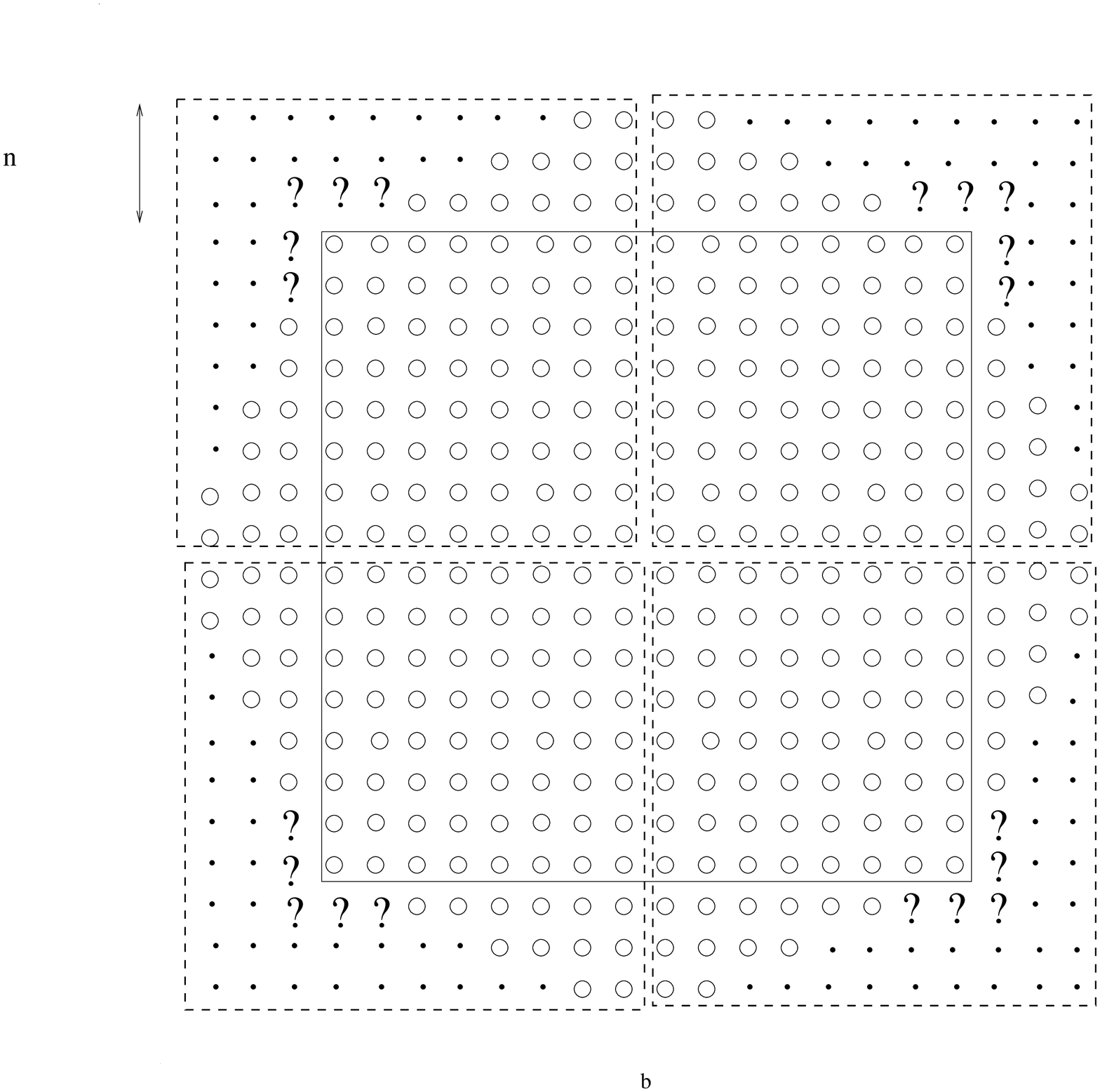}
}
\caption{a) The empty $n\times n$ square. Site $x$ can be emptied
since its SE and SW neighbours are all inside the empty square. b) The 
triangular structure of height $I(n/4)$ which can be emptied above each side. $?$ denotes the $5$ sites on each corner which are not
guaranteed to be emptied (unless a proper condition is fulfilled by the configuration outside the empty $n\times n$ nucleus).}
\label{emptying}
\end{figure}

\begin{figure}
\psfrag{n}[][]{{\small{n/2+I(n/4)}}}
\psfrag{y}[][]{$y$}
\begin{center}
\includegraphics[width=0.45\columnwidth]{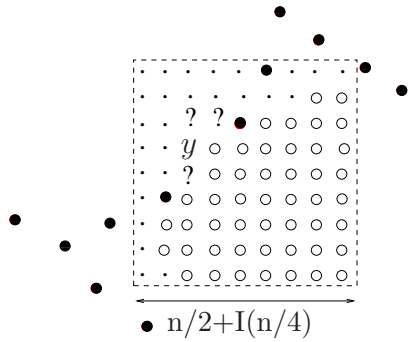}
\end{center}
\caption{Here we depict the top left square in Fig. \ref{emptying} b) and show the necessary condition for $y$ to be frozen: it should belong to a NE-SW cluster spanning the dashed square. Otherwise: if one of its end is free it can be unblocked from there; if they are both blocked by T-junctions with NW-SE clusters these terminate inside the empty region and can be unblocked from there.}
\label{necessary}
\end{figure}

Consider a configuration within which there is an empty square of size
$n\times n$ and focus on the sufficient conditions to empty the next
shell, i.e. to construct an allowed path which empties the $n+2\times
n+2$ square.  The initial vacancies guarantee that we can empty a
centered segment of length $n-4$ external to each side. Consider for
example site $x$ in fig.  \ref{emptying} a), both its SW and SE
couples are inside the empty square: the constraint is satisfied on both diagonals and $x$ can be emptied. 
It is
immediate to check that, even if the rest of the lattice is occupied,
this procedure can be continued until emptying the triangular structure
above each side depicted in Fig. \ref{emptying} b).  This is formed by
centered lines whose length decreases of four sites at each step,
therefore its overall  height
is $I(n/4)$ where $I(x)$ stands for the
integer part of $x$.  In order to empty the $n+2\times n+2$
square, five sites remain to be emptied on each corner: those
indicated by question marks in Fig.\ref{emptying} b).  Consider for
example one of these site on the top left corner, indicated as $y$ in Fig.\ref{necessary}.  The constraint along the
NW-SE diagonal is verified, since both its SE neighbors are inside the
empty square.  Instead, since neither the NE nor the SW couples are
contained in the empty region, $y$ can be blocked along NE-SW diagonal and in this case we cannot empty it.  However, it is possible to derive a necessary condition for $y$ to be
frozen, which gives in turn a sufficient condition to construct an allowed 
(i.e. verifying the constraint at each elementary move)
 path in the configuration space which allows to empty $y$.  
Focus on the square of size $n/2+I(n/4)$
centered on the top left corner and containing half of the triangular
structure associated to the up and left sides, i.e. the square inside
the dashed region in Fig. \ref{emptying}b).  Consider the sublattice
constructed from $y$ inside this square by erasing all sites except
its North-East and South-West neighbors and so on (as we did for the
origin in figure \ref{lattice}).  It can be directly checked that $y$
can be frozen only if it belongs to a NE-SW path, i.e.  a directed
path on this sublattice, which spans the square (see Fig.
\ref{necessary}).
 This is due to the fact that any occupied cluster
along the NW-SE diagonal can be unfrozen if it terminates inside the
empty region ($n\times n$ square plus triangular
region). Therefore, a sufficient condition to construct an allowed path to
empty $y$ (and the other remaining four sites at the top
left corner) is that in the dashed square region there is not a
percolating NE-SW cluster.  Since the size of this region is
proportional to $n$ and $\rho<\rho_{dp}$, such a cluster occurs with 
a very small probability, $\exp(-n/\xi_{\parallel})$ for large $n$
with $\xi_{\parallel}$ the parallel correlation length
 for DP \cite{Hin}. As a consequence  
the probability of emptying $y$  increases at least
exponentially fast to one when the size of the empty nucleus, $n$, is
increased. The same argument
can be applied to the other four sites indicated by question marks and to those on the other three corners of the lattice, 
concluding that the probability
$P(n\to n+2)$ to go from the empty $n\times n$ to the empty $n+2\times
n+2$ nucleus is bounded from below by $
(1-20\exp(-n/\xi_{\parallel}))$.  Thanks to the exponential increase
towards one, the probability that this procedure can be continued up
to infinite size stays strictly positive because it is bounded
from below by $\prod_{i=n}^{\infty}
P(i\to i+2)>0$. Finally, using the $O(L^2)$ different positions for
the initial empty nucleus, we conclude that the probability for
$\Lambda_L$ to be internally spanned converges to one in the
thermodynamic limit at any $\rho<\rho_{dp}$.  This is due to the fact
that, in order to prevent the expansion of a large empty nucleus, we
should require long DP clusters, which are highly improbable below $\rho_{dp}$.

\subsection{Discontinuity of the transition}

Results in Sections \ref{noergo} and \ref{ergo}
prove that an ergodicity
breaking transition occurs for Knights
at $\rho_{dp}$, due to the  
percolation transition of blocked structures. In the present and the following section we show that this transition, 
which we call {\sl jamming percolation},
have  features which are qualitatively different from those of DP
and any conventional percolation transition.
Indeed it is discontinuous, i.e. the density of the frozen clusters 
have a finite jump at the transition (the critical clusters are compact rather than fractal) and their typical size increases
faster than any power law when $\rho\nearrow \rho_{dp}$.

\begin{figure}
\centerline{
\psfrag{a}[][]{a)}
\psfrag{O}[][]{\small{O}}
\psfrag{b}[][]{b)}
\psfrag{c}[][]{c)}
\psfrag{ell412}[][]{$\ell_4/12$}
\psfrag{ell4}[][]{$\ell_4$}
\psfrag{l0}[][]{{{\small{$\ell_0$}}}}
\psfrag{r1}[][]{{{}}}
\psfrag{r2}[][]{}
\psfrag{r3}[][]{}
\psfrag{r4}[][]{}
\psfrag{r5}[][]{}
\psfrag{r6}[][]{}
\psfrag{r7}[][]{}
\psfrag{r8}[][]{}
\includegraphics[width=0.3\columnwidth]{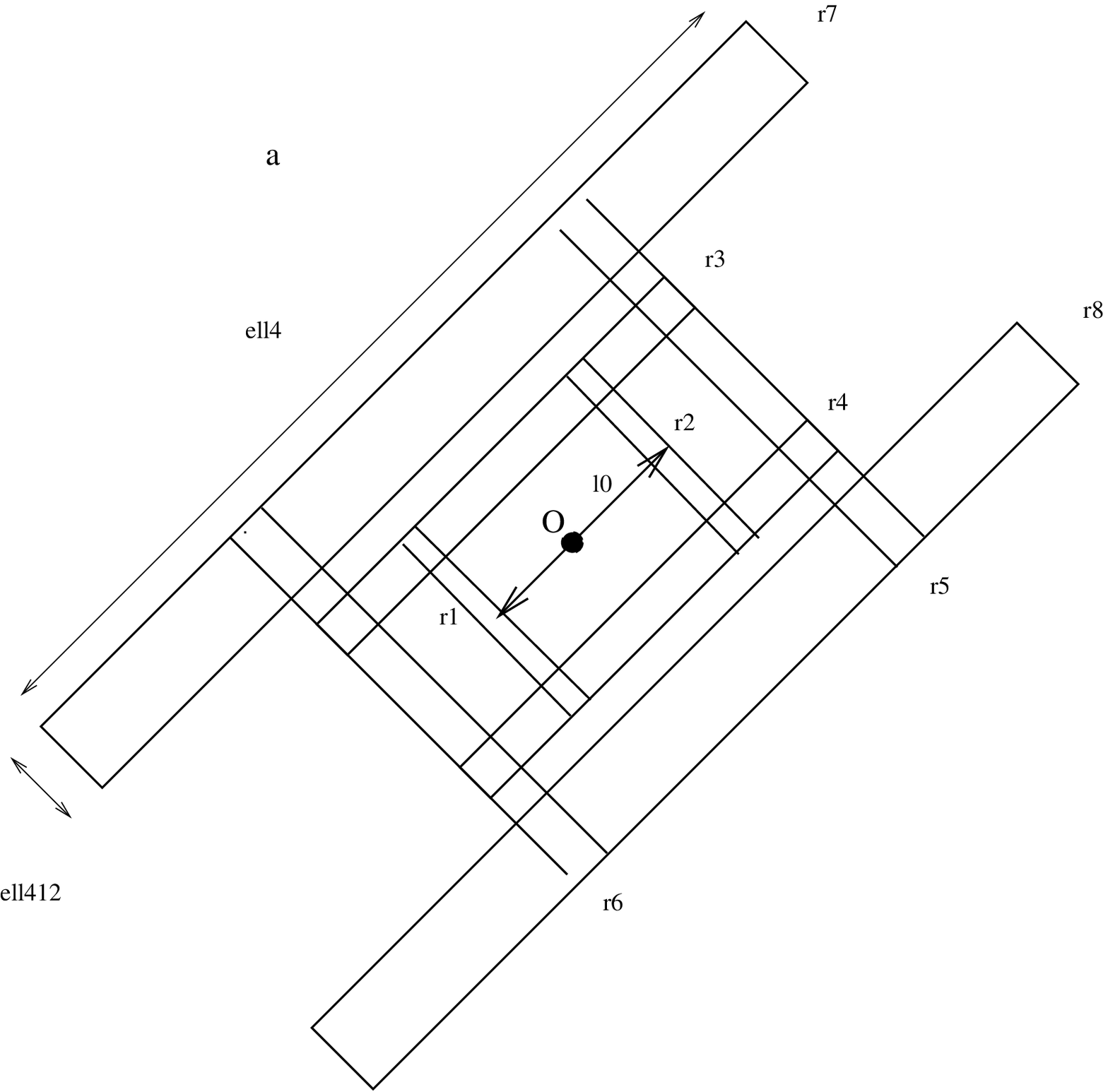}
\hspace{0.3 cm}
\includegraphics[width=0.3\columnwidth]{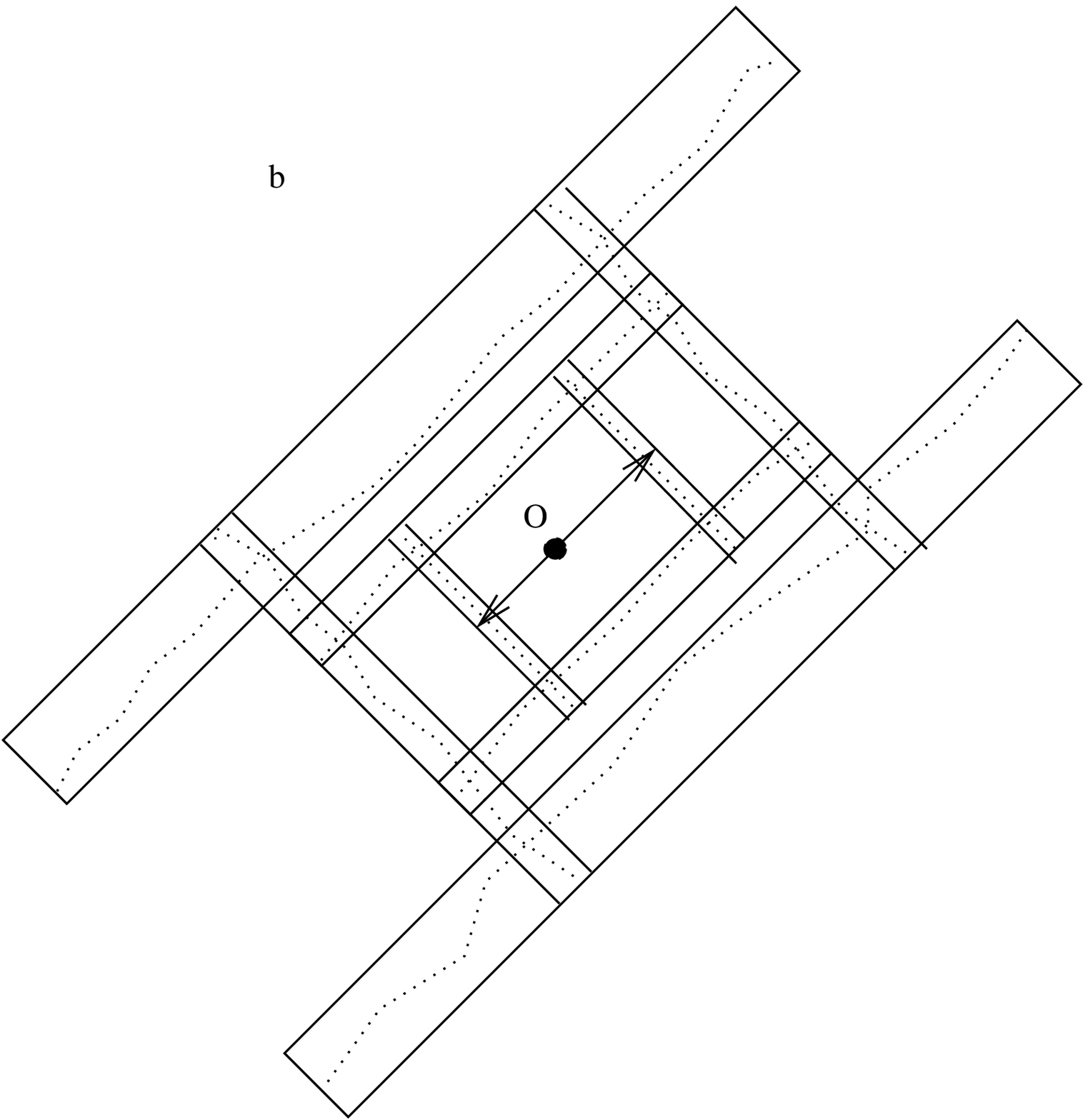}
\hspace{0.3 cm}
\includegraphics[width=0.3\columnwidth]{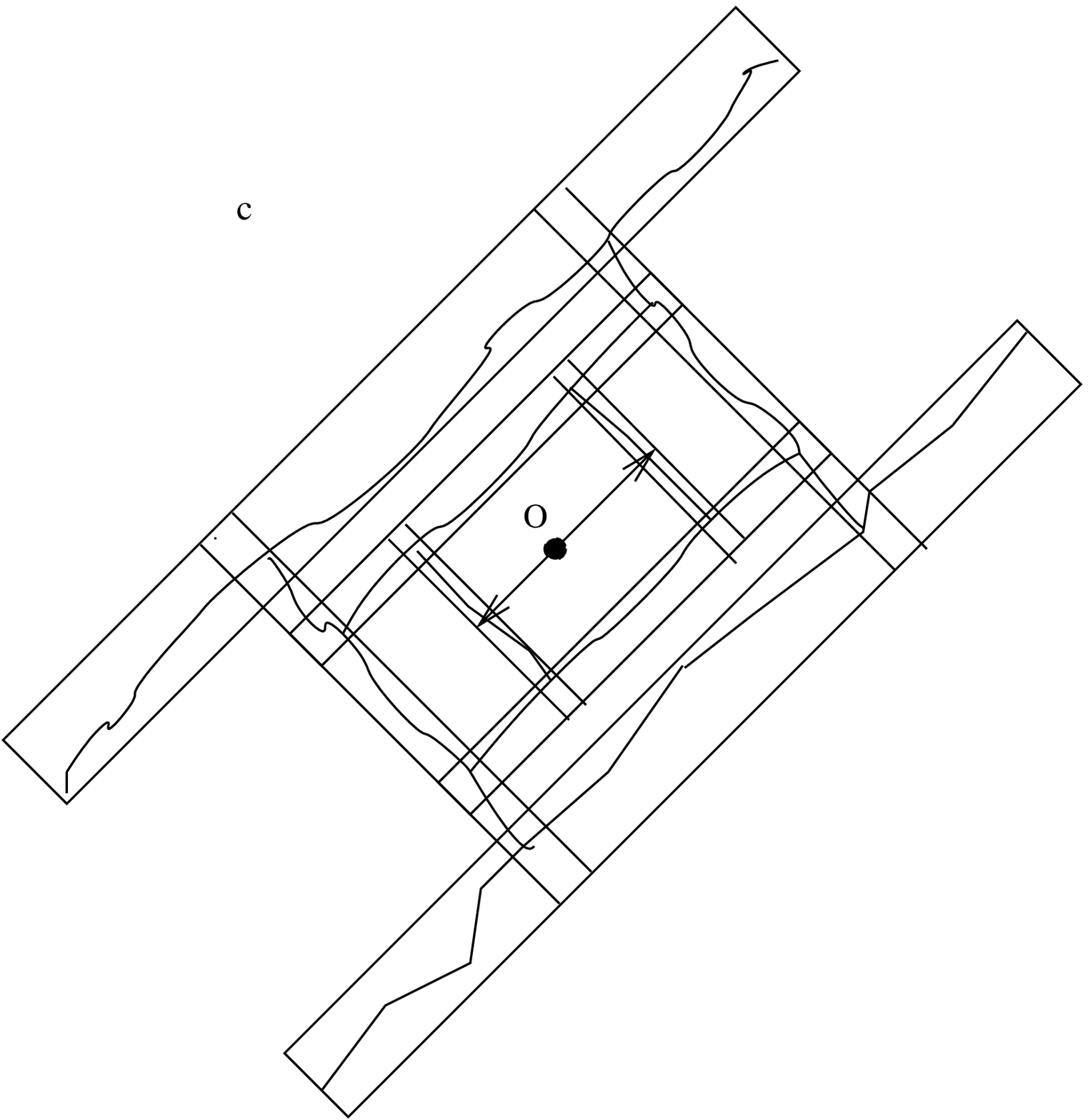}
}
\caption{a) The sequence of intersecting rectangles of increasing size $\ell_i\times 1/12 \ell_i$ describer in the text. b) Dotted non straight line stand for NE-SW (NW-SE) clusters spanning the rectangles
with long side in te NE-SW (NW-SE) direction. c) Frozen structure containing the origin}
\label{disc}
\end{figure}

In order to prove discontinuity 
we construct a set of configurations
which have finite probability 
on the infinite lattice at $\rho_{dp}$ 
and show that the origin is frozen for all these configurations.
We will make use of the blocked structures
containing T-junctions among NE-SW and NW-SE clusters
which have been introduced
in Section \ref{ergo}.
  Consider a configuration in which the origin belongs
to a NE-SW path  of length $\ell_0/2$: this occurs with probability $p_0$. Now
focus on the infinite sequence of pairs of rectangles of increasing
size $\ell_i\times\ell_i/12$ with $\ell_1=\ell_0$,
$\ell_i=2\ell_{i-2}$ and intersecting as in Fig.\ref{disc}.
 If {\sl each} of 
these
rectangles with long side along the NE-SW (NW-SE) diagonal
contains a NE-SW (NW-SE) percolating path (dotted lines in Fig. \ref{disc} b)), the origin is frozen.
This can be directly checked: an infinite backbone  
of mutually blocked particles
constituted by pieces of these paths connected
by T-junctions (continuous line in Fig. \ref{disc} c)) occurs.
Therefore the probability that the origin is frozen, $q(\rho)$, is bounded from below by 

$$q(\rho)>p_o\prod_{i=1,\infty}P(\ell_i)^2$$ where $P(\ell_i)$ is the
probability that a rectangle of size $\ell_i\times 1/12 \ell_i$ with
short side in the transverse direction is spanned by a DP cluster.
This probability converges to one exponentially fast in $\ell_i$ due
to the anisotropy of critical clusters for directed percolation.
Recall that there are a parallel and a transverse length for DP with
different exponents, i.e. a cluster of parallel length $\ell$ has
typically transverse length $\ell^z$. Let us divide the 
$\ell_i\times 1/12 \ell_i$
rectangle into  $\ell_i^{1-z}$ slices of size $\ell_i\times 1/12 \ell_i^{z}$.
For
each slice the probability of having a DP cluster along the parallel
direction at $\rho_{dp}$ is order unity. Thus, the probability of {\sl
not} having a DP cluster in each of the slice is
$1-P(\ell_i)=O[\exp(-c\ell_i^{1-z})]$. From this  result
and above
inequality for $q$, it follows immediately
 $q(\rho_{dp})>0$.

Therefore 
the infinite cluster of jamming
percolation is ``compact'', i.e. of dimension $d$ at the transition.
Note that to obtain discontinuity two ingredients of the constraints
were crucial: 
the existence of two transverse blocking directions
each with an underlying percolation transition;
 the anisotropy of these transitions.
Indeed, anisotropy is necessary to have the increase towards one of 
the probability that above  rectangles are spanned when their size is increased. In turn, this is necessary
to get a finite probability for the construction which freezes the origin.

\subsection{Dynamical correlation length}

In this Section we explain the arguments in \cite{letterTBF} leading
to the result $\log\Xi(\rho)\simeq k (\rho-\rho_{dp})^{-\mu}$ for
$\rho<\rho_{dp}$, where  $\mu=\nu_{\parallel}(1-z)\simeq 0.64$.  
  As already explained, $\Xi(\rho)$ is 
 the crossover length dividing the regime in which the probability
for a  finite region to be internally spanned goes to zero or
to one when $\rho\nearrow\rho_{dp}$, i.e. the length
below which finite size effects are important. 
For jamming percolation it corresponds to the
diverging size of the incipient spanning
clusters  and for Knight models to the
typical size of the region that has to be rearranged to unblock a
given site.
Let us sketch separately the  arguments leading to $\log\Xi(\rho)\geq k_l (\rho-\rho_{dp})^{-\mu}$ and  to 
$\log\Xi(\rho)\leq k_u (\rho-\rho_{dp})^{-\mu}$ with $k_l,k_u$ two positive constants.

\begin{figure}
\psfrag{s}[][]{{\LARGE{$s$}}}
\psfrag{s6}[][]{{\LARGE{$s/6$}}}
\psfrag{A}[][]{{\LARGE{A}}}
\psfrag{B}[][]{{\LARGE{B}}}
\psfrag{C}[][]{{\LARGE{C}}}
\psfrag{D}[][]{{\LARGE{D}}}
\psfrag{E}[][]{{\LARGE{E}}}
\psfrag{F}[][]{{\LARGE{F}}}
\psfrag{G}[][]{{\LARGE{G}}}
\psfrag{H}[][]{{\LARGE{H}}}
\begin{center}
\resizebox{0.4 \hsize}{!}{\includegraphics*{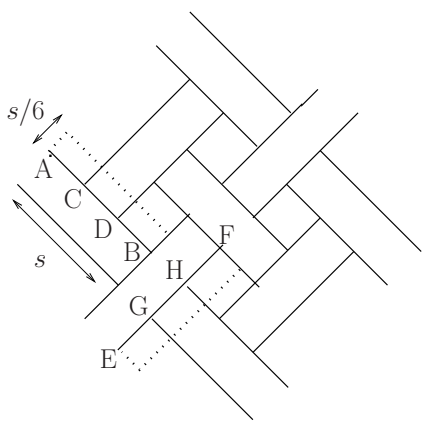}}
\end{center}
\caption{The frozen structure described in the text: continuous lines
  stand for occupied NE-SW or SW-NE clusters (for simplicity we draw them as straight lines,
  recall however that they will in general bend since they have two bending
  directions). Each of these clusters is blocked since it ends in a
  T-junction with a  cluster along the transverse diagonal. 
The dotted rectangle adjacent to cluster AB (EF) are the regions in which this
  cluster can be displaced and
  yet a frozen backbone is preserved. Indeed the T-junctions in C and D (G and
  H) will be displaced but none of them will be disrupted.}
\label{lower}
\end{figure}

To establish the lower bound we construct a set of configurations
 containing a frozen
backbone and we show that the probability of this set
goes to one when $\rho\nearrow\rho_{dp}$ and $L\to\infty$ with $\log
L\leq k_l(\rho-\rho_{dp})^{-\mu}$.  Again, we make use of the
T-junctions described in Section \ref{ergo}.
Consider the set of NE-SW and NW-SE paths of length $s$ intersecting 
as
in Fig. \ref{lower}.

As can be  directly checked, this structure can be emptied only 
starting from its border since each
finite directed path terminates on T-junctions with paths in the 
opposite direction. Therefore, if the construction
 is continued up to the
border and we consider periodic boundary conditions, all sites in 
the structure are frozen. Furthermore, 
a similar frozen backbone exists
also if one or more of these paths is displaced inside an adjacent
rectangular region of size $s\times s/6$, as shown in Fig. 
\ref{lower}. Therefore the probability that the regions is not
internally spanned, $1-R(L,\rho)$, is bounded from below
by the probability that {\sl each} of these rectangles contains
at least one path connecting its short sides.
This leads to
\begin{equation} \label{ineqxx}
R(L,\rho)\leq n(L,s)\exp(-cs^{1-z})
\end{equation}
where $n(L,s)=O(L/s)^2$ is the number of the rectangles contained in the
structure and $\exp(-cs^{1-z})$ is the probability of not having
a DP path in a region $s\times s/6$ as long as $s\leq\xi_{\parallel}$
(see previous section).
 The inequality (\ref{ineqxx}) leads to 
$\lim_{ L\to\infty, \rho\nearrow\rho_{dp}}R(L,\rho)=0$ for 
$\xi_{\parallel}\exp(c\xi_{\parallel}^{1-z})/L\rightarrow \infty$, therefore
$\log\Xi \geq k_l (\rho-\rho_{dp})^{-\mu}$.

On the other hand, in order to establish the upper bound, we
show that for $\log L\geq k_u(\rho-\rho_{dp})^{-\mu}$ there is typically
 an nucleus of vacancies which can be expanded until emptying the
whole lattice. From the results in Section \ref{ergo} it is easy to
see that the probability to expand an empty nucleus to infinity is
dominated by the probability of expanding it up to
$n=\xi_{\parallel}$. Indeed, above this size the probability of an
event which prevents expansion is exponentially suppressed. Therefore,
considering the $O(L/\xi_{\parallel})^2$ possible position for a regions
that it is guaranteed to be emptyable up to size $\xi_{\parallel}$, we
get that the probability that $\Lambda_L$ is internally spanned is
roughly bounded as $R(L,\rho)\geq L^2\delta$ for $L^2\leq \delta$, where $\delta$
is the probability 
that a small empty nucleus can be expanded until size 
$\xi_{\parallel}$.
In  the emptying procedure described in Section \ref{ergo} we 
subsequently increase from $n$ to $n+2$ the size of the empty nucleus.
If we
require instead that on a region of size $O(\xi_{\parallel})$ around each corner there is not a spanning DP cluster, we can expand directly to $\xi_{\parallel}$. The probability of the latter event gives $\delta\geq
\exp(-c\xi_{\parallel}^{1-z})$ (see \cite{letterTBF}). Therefore, we get
the desired upper bound on $\log\Xi$ 
that has the same scaling as the lower bound.

\subsection{Numerical Results}

In the following we present numerical results on the percolation of
blocked structures that support our theoretical findings. Starting 
from an initial configuration
we  sequentially pick away all particles that are not
blocked, so that the final configuration contains the backbones of
forever blocked particles.
\begin{figure}
\centerline{
{\includegraphics[width=0.9\columnwidth]{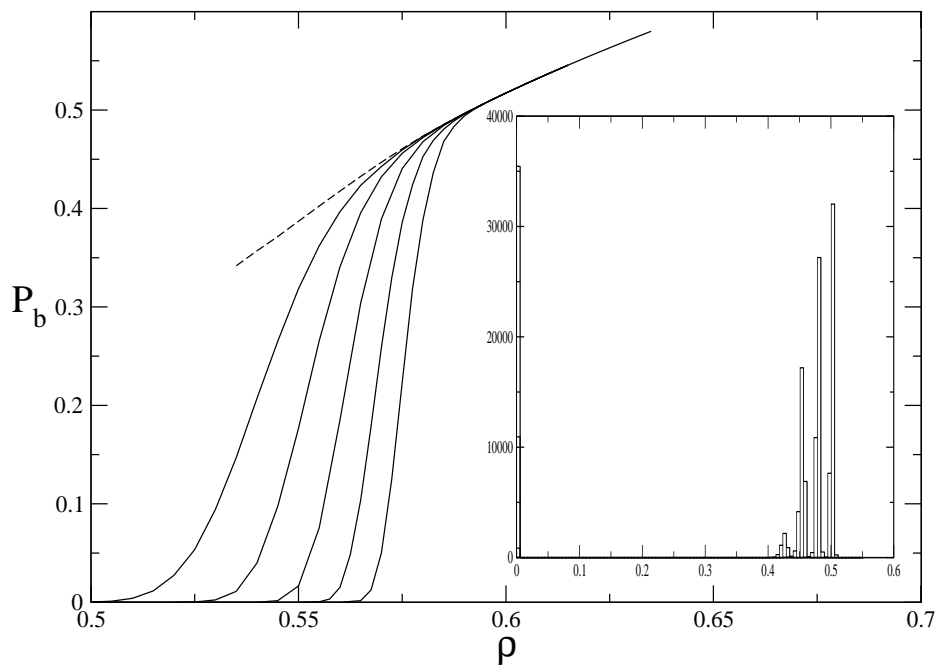}}}
\caption{Probability that a site belongs to the infinite blocked
  spanning cluster as a function of the initial density for system
  sizes $N=100^2,200^2,400^2,800^2,1600^2$ averaged over $40000$
  initial configurations (from left to right respectively). 
 Inset: histograms of the number of initial
  configurations (y axis) leading to a fraction of sites $\rho_b$
  (x-axis) for $N=800^2$, $40000$ initial configurations and initial
  density $\rho=0.5775,0.5825,0.5875,0.5925$ (when 
increasing the initial
  density the right peak moves to the right).}
\label{pb}
\end{figure}
In Fig. \ref{pb} we show the probability that a site
belongs to the infinite spanning cluster for lattice sizes
$N=100^2,200^2,400^2,800^2,1600^2$ averaged over $40000$
samples. Though the transition takes place at the critical density 
of DP $\rho_{dp}\simeq 0.701$, as we
have proved, finite size effects appear already at $0.53-0.55$.

Indeed, the probability that there {\it exists} a frozen cluster 
is substantial for
$\rho$ fifteen percent below $\rho_c$ even in our largest systems,
($L=1600$): it is thus hard to study the asymptotic critical behavior (see
\cite{Dawson} for an analogous problem in the context of bootstrap
percolation).   But in a slightly different model one can get closer to the
transition \cite{TBF2long}: these data are consistent with the  predicted $\ln\Xi \sim (\rho_c-\rho)^{-\mu}$ with $\mu\cong 0.64$, but the small range of $\ln L$ available makes the uncertainties in $\mu$ large.  

Another theoretical finding confirmed
by numerics is the first order character of the transition. Indeed the
histograms of the density of blocked structures, $\rho_b$,
clearly show a well defined two peak shape as in
usual first order transitions  (see inset of
Fig. \ref{pb}).  The peak at a non zero density
is clearly distinct from the one at zero density and moves to the right
by increasing the density of particles in the initial configuration, thus
showing that asymptotically the transition is first order. Furthermore
there are no finite size effects on the peak position but only on its
weight.  Thus, the continuous curves $P_b (\rho)$ in Fig. \ref{pb} are
the average of two curves: the trivial one, $P_b=0$, corresponding to
the left peak and the dotted one corresponding to the right
peak. By changing the system size the weights of the two peaks
change and so the continuous averaged $P_b(\rho)$ shifts to the right.

\begin{figure}
\centerline{\rotatebox{-90}{\includegraphics[width=0.6\columnwidth]{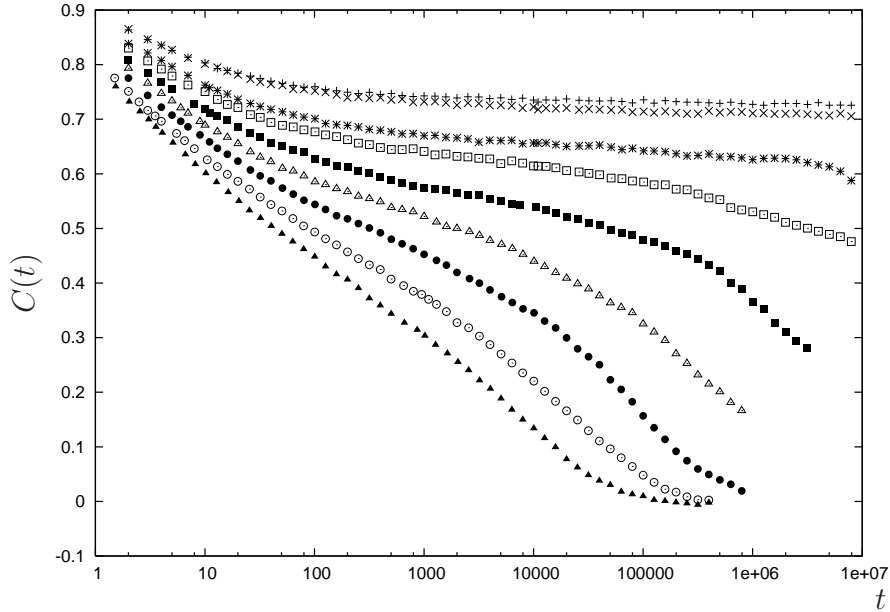}}}
\caption{Density density correlation as a function of time (MonteCarlo
  steps) for a $100\times 100$ lattice. From bottom to top,
  $\rho=0.49,0.50,0.51,0.52,0.53,0.54,0.55,0.56,0.57$.}
\label{corr}
\end{figure}

Finally, let's focus on the predictions for the dynamics. 
First, as already discussed, an
ergodicity breaking transition occurs at $\rho=\rho_{dp}$.  Furthermore,
the first-order character of the percolation of blocked structures
implies a discontinuous jump of
$q_{EA}(\rho)=\lim_{t\to\infty}<\eta_x(t)\eta_x(0)>_c$. This is the plateau 
of the correlation function, see Fig.\ref{corr} and is the analog of the
Edwards-Anderson parameter in spin-glasses. 
\psfrag{C(t)}[][]{$C(t)$}
\psfrag{t}[][]{$t$}\
In fact for $\rho<\rho_c$, ergodicity
implies that $q_{EA}(\rho)=0$ since the system always relaxes to
equilibrium. Instead for $\rho\ge \rho_c$, by separating the
contribution from sites which are occupied and blocked forever and the
remaining sites, one can show that $q_{EA}(\rho)\geq c(\rho)q(\rho)$ 
\cite{letterTBF,longTB}, where $c(\rho)$ is a strictly positive 
for all $\rho<1$ and $q(\rho)$, as defined
previously, is the probability the the origin is frozen. Therefore,
$q_{EA}(\rho)>0$, for $\rho\ge\rho_{dp}$.  

Furthermore the relaxation timescale, $\tau$, 
diverges also very fast at the transition as discussed previously. 
As a check, we performed numerical simulations of standard Montecarlo
dynamics. The results for 
$<\eta_x(t)\eta_x(0)>_c$ are plotted in Fig. \ref{corr} for system
size $N=100^2$. The curves clearly shows a developing (discontinuous)
plateau which becomes infinite after a certain (size dependent)
density. However, the initial curves until $\rho=0.53$ have no finite
size effects. This shows that the developing of a discontinuous
plateau and the increasing of the timescale start far from the
critical density, in agreement with an essential singularity for the
laws for $\Xi,\tau$ (for a power law divergence $\tau$ should diverge 
much closer to $\rho_{c}$). 
Note that, because of the very severe finite size effects, 
one needs much larger system sizes  (and much larger timescales) 
in order to measure successfully the exponent of the essential 
singularity. We leave that for future study.

\section{Conclusion}
\label{conclu}
As discussed in this work cooperative KCMs display a remarkable 
physical behavior. In particular, Knights model undergoes a purely 
dynamical phase transition due to a jamming percolation : a giant
blocked cluster appears at $\rho_c<1$. 
This new type of transition has features 
that are very different from usual (first and second order) 
phase transitions and that  are similar to the ones indeed expected for 
glass-jamming transitions. In particular the fraction of jammed particles 
is discontinuous (as for first order phase transition) 
although time and lengthscales diverge (as for second order phase transitions).
Furthermore the relaxation time diverges with a Vogel-Fulcher like 
form, i.e. much faster than a power law. 

The extension and the universality of our results are fundamental open 
questions. In three dimensions, two natural generalizations of our jamming percolation exist: one composed of DP clusters --- which should slow down as a double exponential of $(\rho-\rho_c)^{{-\overline{\mu}}}$ ---  and the other of directed sheet-like structures which will have exponential slowing down like we have found in 2D.  The key ingredients are  kinetic constraints that enable huge jammed clusters to form out of small objects without these becoming much more common or much larger. A very important issue is whether the mechanism that we devise to create a jamming percolation, which is based on interacting DP clusters,
is the only possible one or there are others that may lead to a very similar
dynamical transition, see \cite{Schwarz}.  

For the future, the connection between our results and the jamming transition found for continuum particle systems \cite{colloids} needs exploring. 
One should analyze the effects of
constraint-violating processes occurring with a very low rate: these are 
certainly present for molecular liquids undergoing a glass transition. 
Similarly, it would be very important to generalize our results to systems 
of particles in the continuum. Finally, it would be very interesting to 
compare the geometrical and statistical properties of jamming percolation 
clusters to the ones that can be measured in experiments in 
colloidal and granular systems.\\

{\bf Acknowledgments}\\
Many of the results that we presented were obtained with D.S.Fisher. 
We are very happy to thank him for the long-standing collaboration
on glass-jamming transitions. We thank also M. Sellitto for 
having investigated with us the jamming transition on the Bethe 
lattice summarized in Section V. We are grateful to L. Berthier
for helpful discussions and collaboration on related subjects.  
GB is partially supported by the European Community's Human Potential Program
contracts HPRN-CT-2002-00307 (DYGLAGEMEM).


\begin{thebibliography}{99}
\bibitem{liquids}
P.G.De Benedetti and F.H.Stillinger {\sl Nature} {\bf  410}, 267 (2001).
\bibitem{colloids}
ER.Weeks et al.
{\sl Science}
{\bf  287}, 627 (2000);
V.Trappe et al.
{\sl Nature} {\bf 411}, (2001)
722.
\bibitem{Dau}G.Marty, O.Dauchot {\sl Phys.Rev.Lett} {\bf 94} 015701 (2005) 
\bibitem{hetexp} M.A.Ediger {\sl Ann.Rev.Phys.Chem.} {\bf 51}, 99 (2000); E.Vidal-Russel, N.E.Israeloff {\sl Nature} {\bf 408}, 695 (2000); L.A.Deschenes, D.A.Vande Bout {\sl Science} {\bf 292} 255 (2001); L.Berthier et al. {\sl Science} {\bf 310} 1797 (2005); C.Bennemann et al.
{\sl Nature}, {\bf 399}, 246 (1999);O.Dauchot, G.Marty, G.Biroli, {\sl Phys.Rev.Lett} {\bf 95}, 265701 (2005).
\bibitem{KTW}T.R.Kirkpatrick, D.Thirumalai, {\sl Phys.Rev.Lett} 
{\bf 58}, (1987) 2091; T.R.Kirkpatrick, D.Thirumalai, P.G.Wolynes {\sl Phys.Rev.A} {\bf 40}, (1989) 1045.
\bibitem{BoCuKuMe}  Bouchaud J-P, Cugliandolo L F, Kurchan J and M{\'e}zard M, in {\it
    Spin-Glasses and Random Fields}, edited by Young A P (World
  Scientific, 1997).
\bibitem{Wolynes} X. Xia and P.G. Wolynes, PNAS {\bf 97} 2990 (2000) and refs therein.
\bibitem{BB} J.-P. Bouchaud and G. Biroli, J. Chem. Phys. {\bf 121}, 7347 (2004).

\bibitem{Franz}S.Franz {\sl Europhys.Lett.} {\bf 73} 492 (2006)
\bibitem{WolynesDzero} M. Dzero, J. Schmalian, P. G. Wolynes, cond-mat/0502011.  
\bibitem{Moore}M.A.Moore cond-mat/060241
\bibitem{J}J. Jackle J. Phys. Cond. Matter {\bf 14}, (2002) 1423;
\bibitem{RS} F. Ritort, P. Sollich {\sl Adv. in Phys.} {\bf 52} (2003), 219.
\bibitem{FA} G.H.Fredrickson, H.C.Andersen {\sl Phys.Rev.Lett.} {\bf 53}, 1244 (1984); {\sl J.Chem.Phys.} {\bf 84}, 5822 (1985)
\bibitem{letterTBF} C.Toninelli,G.Biroli, D.S.Fisher{\sl Phys.Rev.Lett} {\bf 96}, 035702 (2006).  
\bibitem{JE} J.Jackle, S.Eisinger {\sl Z.Phys.B: Cond.Mat.} {\bf 84}, 115 (1991)\bibitem{KA}
W. Kob and H.C. Andersen, {\sl Phys.Rev.E} {\bf 48} (1993) 4364.
\bibitem{SE} P.Sollich, M.Evans {\sl Phys.Rev.Lett} {\bf 83} 3238 (1999)
\bibitem{AD} D.Aldous, P.Diaconis {\sl J.Stat.Phys.} {\bf 107} 845 (2002)
\bibitem{GC}   J.P.Garrahan, D.Chandler {\sl Phys. Rev.Lett} {\bf 89} 035704 (2002)
\bibitem{WBG} S.Whitelam, L.Berthier, J.P.Garrahan, {\sl Phys. Rev.Lett} {\bf 92}, 185705 (2004); {\sl Phys.Rev.E} {\bf 71}, 026128 (2005)
\bibitem{JMS} R.Jack, P.Mayer, P.Sollich cond-mat/0601529 
\bibitem{CMRT} N.Cancrini, F.Martinelli, C.Roberto, C.Toninelli cond-mat/0603745
\bibitem{BG} L.Berthier, J.P.Garrahan {\sl J.Chem.Phys.} {\bf 119}, 4367 (2003)
\bibitem{GCPNAS} J.P. Garrahan, D. Chandler, Proc. Natl. Acad. Sci. {\bf 100}, 9710 (2003).
\bibitem{R} J.Reiter {\sl J.Chem.Phys} {\bf 95}, 544 (1991)
\bibitem{FB} G.H.Fredrickson, S.A.Brawer {\sl J.Chem.Phys} {\bf 84} 3351 (1986)
\bibitem{ES} M.Einax, M.Schulz {\sl J.Chem.Phys} {\bf 115} 2282 (2001)
\bibitem{GPG} I.S.Graham, L.Pich{\'e}, M.Grant, {\sl J.Phys.Cond.Matt}, {\bf 5}, 6491 (1993); {\sl Phys.Rev.E}, {\bf 55}, 2132 (1997)
\bibitem{H} P.Harrowell, {\sl Phys.Rev.E} {\bf 48}, 4359 (1993)
\bibitem{F}G.H.Fredrickson {\sl Ann.N.Y.Acad.Sci}, {\bf 484}, 185 (1986)
\bibitem{BH}S.Butler, P.Harrowell, {\sl J.Chem.Phys.} {\bf 95}, 4466 (1991)

\bibitem{TBF2long} C. Toninelli, G. Biroli and D.S. Fisher in preparation.
\bibitem{Berthier} L. Berthier, G. Biroli, C. Toninelli in preparation.
\bibitem{RJM} J.Reiter, F.Mauch, J.Jackle, {\sl Physica A}, {\bf 184}, 493 (1992)
\bibitem{noiKA} C.Toninelli,G.Biroli, D.S.Fisher {\sl Phys.Rev.Lett.} {\bf 92} 185504 (2004); {\sl J.Stat.Phys.} {\bf 120}, 167 (2005)
\bibitem{longTB} C.Toninelli, G.Biroli cond-mat/0512335
\bibitem{Kob} W. Kob, in "Slow relaxations and nonequilibrium dynamics in condensed matter", vol. Session LXXVII of Les Houches Summer School (ed. J.-L. Barrat, M. Feigelman and J. Kurchan), published by EDP Sciences and Springer.
\bibitem{WW} E.R.Weeks, D.A.Weitz {\sl Phys.Rev.Lett.} {\bf 89} 095704 (2002); {\sl Chem.Physics} {\bf 284}, 361 (2002)
\bibitem{BT} L.Bertini, C.Toninelli {\sl J.Stat.Phys.}, {\bf 117}, 549-580  (2004) 
\bibitem{TB}  C.Toninelli, G.Biroli 
{\sl J.Stat.Phys.}, {\bf 117}, 27-54  (2004)
\bibitem{spohn} H.Spohn, {\sl Large scale dynamics of interacting particles} Berlin Springer (1991)

\bibitem{Adler} J.Adler {\sl Physica A} {\bf 171 } (1991) 435

\bibitem{AL}
M. Aizenmann, J.L. Lebowitz, {\sl J.Phys.A}
{\bf 21} (1988) 3801.
\bibitem{k-core} Pittel B, Spencer J and Wormald N, {\it J. Comb. Th.
    B} {\bf 67} (1996) 111.
\bibitem{Sch} 
R.H.Schonmann, {\sl Ann. of Probab.},{\bf 20} 
(1992), 174-193 
\bibitem{librolandim} C. Kipnis, C. Landim, {\it Scaling Limits of Interacting Particle Systems},
Grundlheren der Mathematischen Wissenschaften 320, Springer-Verlag, Berlin, New York, 1999.
\bibitem{LT} C.Landim, C.Toninelli in preparation
\bibitem{S} H.Spohn {\sl J.Stat.Phys.} {\bf 59} (19990) 1227; {\sl Physica A} {\bf 163},
(1990) 134.
\bibitem{jorgemauro} J.Kurchan, L.Peliti, M.Sellitto {\sl Europhysics Lett.} {\bf 39} (1997) 365
\bibitem{Sellitto} M. Sellitto, G. Biroli and
C. Toninelli. Europhys. Lett. {\bf  69} (4), 496 (2005).
\bibitem{ChLeRe} Chalupa J, Leath P L and Reich R, {\it J. Phys. C:
    Solid State Phys.} {\bf 12} (1979) L31.
\bibitem{Mendes} S. N. Dorogovtsev, A. V. Goltsev, and J. F. F. Mendes 
Phys. Rev. Lett. {\bf  96}, 040601 (2006). 
\bibitem{Schwarz} J.M. Schwarz, A.J. Liu, L.Q. Chayes,
Europhys. Lett. {\bf  73}, 560 (2006).
\bibitem{Montanari}  Abdelaziz Amraoui, Andrea Montanari, Tom Richardson, Rudiger Urbanke, 42nd Allerton Conference on Communication, Control and Computing , cs.IT/0410019. A. Dembo and A. Montanari in preparation.
\bibitem{Hin} H.Hinrichsen {\sl Adv. in Phys} {\bf 49}
815 (2000)
\bibitem{FrPa} Franz S and Parisi G, {\it J. Phys.: Cond. Matter} {\bf
    12} (2000) 6335
\bibitem{Dawson} 
P.De Gregorio, A.Lawlor, P.Bradley, K.A.Dawson Phys.Rev.Lett. {\bf 93}
 (2004) 025501.


\end{thebibliography}
\end{document}